\documentclass[preprint2]{proto}
\usepackage{times}
\usepackage{epsfig}
\newcommand{\refs}{\par\noindent\hangindent=1pc\hangafter=1}
\begin{document}

\title{\textbf{\LARGE Gravitational Instabilities in Gaseous Protoplanetary Disks
              and Implications for Giant Planet Formation}}

\author {\textbf{\large Richard H. Durisen}}
\affil{\small\em Indiana University}

\author {\textbf{\large Alan P. Boss}}
\affil{\small\em Carnegie Institution of Washington}

\author {\textbf{\large Lucio Mayer}}
\affil{\small\em Eidgen\"ossische Technische Hochschule Z\"urich}

\author {\textbf{\large Andrew F. Nelson}}
\affil{\small\em Los Alamos National Laboratory}

\author {\textbf{\large Thomas Quinn}}
\affil{\small\em University of Washington}

\author {\textbf{\large W. K. M. Rice}}
\affil{\small\em University of Edinburgh}

\begin{abstract}
\baselineskip = 11pt
\leftskip = 0.65in
\rightskip = 0.65in
\parindent=1pc
{\small Protoplanetary gas disks are likely to experience gravitational
instabilites (GI's) during some phase of their evolution. Density 
perturbations in an unstable disk grow on a dynamic time scale into 
spiral arms that produce efficient outward transfer of angular momentum 
and inward transfer of mass through gravitational torques. In a cool
disk with rapid enough cooling, the spiral arms in an unstable disk form 
self-gravitating clumps. Whether gas giant protoplanets can form by 
such a disk instability process is the primary question addressed by this 
review. We discuss the wide range of calculations undertaken by 
ourselves and others using various numerical techniques, and we report 
preliminary results from a large multi-code collaboration. Additional 
topics include --  triggering mechanisms for GI's, disk heating and cooling, 
orbital survival of dense clumps, interactions of solids with GI-driven 
waves and shocks, and hybrid scenarios where GI's facilitate core 
accretion. The review ends with a discussion of how well disk instability 
and core accretion fare in meeting observational constraints.
\\~\\~\\~}
 
\end{abstract}  

\section{\textbf{INTRODUCTION}}

Gravitational instabilities (GI's) can occur in any region of a gas disk
that becomes sufficiently cool or develops a high enough surface 
density. In the nonlinear regime, GI's can produce local and global
spiral waves, self-gravitating turbulence, mass and angular momentum
transport, and disk fragmentation into dense clumps and substructure.
The particular emphasis of this review article is the possibility 
({\it Kuiper}, 1951; {\it Cameron}, 1978), recently revived by {\it Boss} 
(1997, 1998a), that the dense clumps in a disk fragmented by GI's may 
become self-gravitating precursors to gas giant planets. This particular 
idea for gas giant planet formation has come to be known as the 
{\sl disk instability} theory. We provide here a thorough review 
of the physics of GI's as currently understood through a wide variety of 
techniques and offer tutorials on 
key issues of physics and methodology. The authors assembled 
for this paper were deliberately chosen to represent the full range of 
views on the subject. Although we disagree about some aspects of GI's 
and about some interpretations of available results, we have labored 
hard to present a fair and balanced picture. Other recent reviews of this
subject include {\it Boss} (2002c), {\it Durisen et al.} (2003), and 
{\it Durisen} (2006).

\section{\textbf{PHYSICS OF GI's}}

\noindent
\textbf {2.1 Linear Regime}
\bigskip

The parameter that determines whether GI's occur in thin gas disks is 
\begin{equation}\label{eq:Toomre-Q}
Q = c_s\kappa/\pi G\Sigma,
\end{equation}
where $c_s$ is the sound speed, $\kappa$ is the epicyclic frequency at
which a fluid element oscillates when perturbed from circular motion,
$G$ is the gravitational constant, and $\Sigma$ is the surface
density. In a nearly Keplerian disk, $\kappa \approx$ the rotational
angular speed $\Omega$. For axisymmetric (ring-like) disturbances,
disks are stable when $Q > 1$ ({\it Toomre}, 1964). At high
$Q$-values, pressure, represented by $c_s$ in (1), stabilizes short
wavelengths, and rotation, represented by $\kappa$, stabilizes long
wavelengths. The most unstable wavelength when $Q < 1$ is given by
$\lambda_{m} \approx 2\pi^2G\Sigma/\kappa^2$.

Modern numerical simulations, beginning with {\it Papaloizou and
Savonije} (1991), show that nonaxisymmetric disturbances, which grow
as multi-armed spirals, become unstable for $Q \lesssim$ 1.5. Because 
the instability is both linear and dynamic, small perturbations grow
exponentially on the time scale of a rotation period $P_{rot} =
2\pi/\Omega$. The multi-arm spiral waves that grow have a
predominantly trailing pattern, and several modes can appear
simultaneously ({\it Boss}, 1998a; {\it Laughlin et al.}, 1998; {\it Nelson et al.},
1998; {\it Pickett et al.}, 1998). Although
the star does become displaced from the system center of mass
({\it Rice et al.}, 2003a) and one-armed structures can occur (see
Fig. 1 of {\it Cai et al.}, 2006), one-armed modes do not play the
dominant role predicted by {\it Adams et al.} (1989) and 
{\it Shu et al.} (1990).  

\bigskip
\noindent
\textbf {2.2 Nonlinear Regime}
\bigskip

Numerical simulations (see also Sections 3 and 4) show that, as GI's
emerge from the linear regime, they may either saturate at nonlinear
amplitude or fragment the disk. Two major effects control or limit
the outcome -- disk thermodynamics and nonlinear mode
coupling. At this point, the disks also develop large surface
distortions.

\begin{figure}[ht]
\centerline{\epsfig{figure=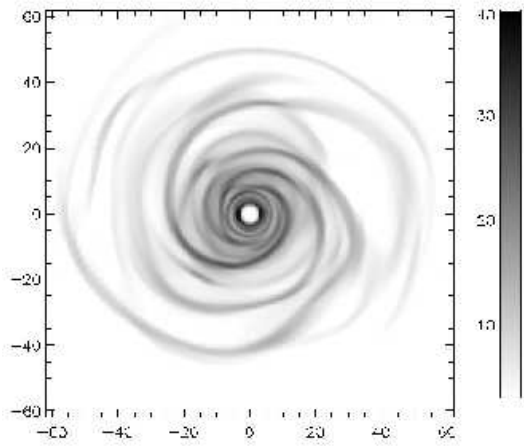,width=75.0mm}}
\caption{\small  
Greyscale of effective temperature $T_{eff}$ in degrees Kelvin 
for a face-on GI-active disk in an asymptotic state of thermal 
self-regulation. This figure is for the {\it Mej\'ia et al.} (2005) evolution 
of a 0.07 $M_{\odot}$ disk around a 0.5  $M_{\odot}$ star
with $t_{cool} = 1$ outer rotation period at 4,500 yr. The 
frame is 120 AU on a side.} 
\label{teff}
\end{figure}

Disk Thermodynamics. 
As the spiral waves grow, they can steepen into
shocks that produce strong localized heating  ({\it Pickett et al.},
1998, 2000a; {\it Nelson et al.}, 2000). Gas is also heated by
compression and through net mass transport due to gravitational
torques. The ultimate source of GI heating is work done by gravity.
What happens next depends on whether a balance can be reached between
heating and the loss of disk thermal energy by radiative or convective
cooling. The notion of a balance of heating and cooling in the
nonlinear regime was described as early as 1965 by {\it Goldreich
and Lynden-Bell} and has been used as a basis for proposing
$\alpha$-treatments for GI-active disks ({\it Paczy\'nski}, 1978; {\it
Lin and Pringle}, 1987). For slow to moderate cooling rates, numerical
experiments, such as in Fig. \ref{teff}, verify that thermal self-regulation 
of GI's can be achieved
({\it Tomley et al.}, 1991; {\it Pickett et al.}, 1998, 2000a, 2003;
{\it Nelson et al.}, 2000; {\it Gammie}, 2001; {\it Boss}, 2003; {\it
Rice et al.}, 2003b; {\it Lodato and Rice}, 2004, 2005; {\it Mej\'ia
et al.} 2005; {\it Cai et al.}, 2006). $Q$ then hovers
near the instability limit, and the nonlinear amplitude is controlled
by the cooling rate.

Nonlinear Mode Coupling. Using second and third-order governing
equations for spiral modes and comparing their results with a full
nonlinear hydrodynamics treatment, {\it Laughlin et al.} (1997, 1998)
studied nonlinear mode coupling in the most detail. Even if only
a single mode initially emerges from the linear regime, power is
quickly distributed over modes with a wide variety of wavelengths and
number of arms, resulting in a self-gravitating turbulence that
permeates the disk. In this {\sl gravitoturbulence}, gravitational
torques and even Reynold's stresses may be important over a wide range
of scales ({\it Nelson et al.}, 1998; {\it Gammie}, 2001; {\it Lodato
and Rice}, 2004; {\it Mej\'ia et al.}, 2005). 

Surface Distortions. As emphasized by {\it Pickett et al.} (1998,
2000, 2003), the vertical structure of the disk plays a crucial role,
both for cooling and for essential aspects of the dynamics. 
There appears to be a relationship between GI spiral modes and the
surface or f-modes of stratified disks ({\it Pickett et al.}, 1996;
{\it Lubow and Ogilvie}, 1998). As a result, except for isothermal
disks, GI's tend to have large amplitudes at the surface of the disk.
Shock heating in the GI spirals can also disrupt vertical
hydrostatic equilibrium, leading to rapid vertical expansions that
resemble hydraulic jumps ({\it Boley et al.}, 2005; {\it Boley and 
Durisen}, 2006). The resulting spiral corrugations can produce 
observable effects (e.g., masers, {\it Durisen et al.}, 2001).

\bigskip
\noindent
\textbf {2.3 Heating and Cooling}
\bigskip

Protoplanetary disks are expected to be moderately thin, with $H/r
\sim 0.05-0.1$, where $H$ is the vertical scale height and $r$ is the
distance from the star. For hydrostatic equilibrium in the vertical
direction, $H \approx c_s/\Omega$. The ratio of disk internal energy
to disk binding energy $\sim c{_s}^2/(r\Omega)^2 \sim (H/r)^2$ is then
$\lesssim 1$\%. As growing modes become nonlinear, they tap the
enormous store of gravitational energy in the disk. Simulation of the
disk energy budget must be done accurately and include all relevant
effects, because it is the disk temperature, through $c_s$ in 
equation \ref{eq:Toomre-Q}, that determines whether the disk becomes
or remains unstable, once the central mass, which governs most of
$\kappa$, and the disk mass distribution $\Sigma$ have been specified.

\noindent
{\it 2.3.1 Cooling}

There have been three approaches to cooling -- make
simple assumptions about the equation of state (EOS), 
include idealized cooling characterized by a cooling time, 
or treat radiative cooling using realistic opacities. 

EOS. This approach has been used to study mode coupling (e.g., {\it
Laughlin et al.}. 1998) and to examine disk fragmentation in the
limits of isentropic and isothermal behavior ({e.g., {\it Boss}, 1998a,
2000; {\it Nelson et al.}, 1998; {\it Pickett et al.}, 1998, 2003;
{\it Mayer et al.}, 2004). Isothermal evolution of a disk, where the disk 
temperature distribution is held fixed in space or when following fluid
elements, effectively assumes rapid loss of energy produced by shocks
and $PdV$ work. Isentropic evolution, where specific entropy is held 
fixed instead of temperature, is a more moderate assumption but is 
still lossy because it ignores entropy generation in shocks 
({\it Pickett et al.}, 1998, 2000a). Due to the energy loss, we do not 
refer to such calculations as adiabatic. Here, we restrict 
{\sl adiabatic evolution} to mean cases where the fluid is treated as an 
ideal gas with shock heating included via an artificial viscosity term 
in the internal energy equation but no radiative cooling. 
Such calculations are adiabatic in the sense that there is no 
energy loss by the system. Examples include a simulation in 
{\it Pickett et al.} (1998) and simulations in {\it Mayer et al.} 
(2002, 2004). {\it Mayer et al.} use adiabatic evolution throughout 
some simulations, but, in others that are started with a locally 
isothermal EOS, they switch to adiabatic evolution as the disk 
approaches fragmentation.

Simple Cooling Laws. Better experimental control over energy loss
is obtained by adopting simple cooling rates per unit
volume $\Lambda = \epsilon/t_{cool}$, where $\epsilon$ is the internal
energy per unit volume. The $t_{cool}$ is specified either as a fixed
fraction of the local disk rotation period $P_{rot}$, usually by setting
$t_{cool}\Omega =$ constant ({\it Gammie}, 2001; {\it Rice et al.},
2003b; {\it Mayer et al.}, 2004b, 2005) or $t_{cool} =$ constant 
everywhere ({\it Pickett et al.}, 2003; {\it Mej\'ia et al.}, 2005). 
In the {\it Mayer et al.} work, the cooling is turned off in dense regions
to simulate high optical depth. Regardless of $t_{cool}$ 
prescription, the amplitude of the GI's in the {\sl asymptotic state} 
(see Fig. \ref{teff}), achieved when heating and
cooling are balanced, increases as $t_{cool}$ decreases. In addition to
elucidating the general physics of GI's, such studies address whether
GI's are intrinsically a local or global phenomenon ({\it Laughlin and
R\'o\.zyczka} 1996; {\it Balbus and Papaloizou}, 1999) and whether they
can be properly modeled by a simple $\alpha$ prescription.  When
$t_{cool}$ is globally constant, the transport induced by GI's is
global with high mass inflow rates 
({\it Mej\'ia et al.}, 2005; {\it Michael et al.}, in preparation); when
$t_{cool}\Omega$ is constant, transport is local, except for thick or
very massive disks, and the inflow rates are well
characterized by a constant $\alpha$ ({\it Gammie}, 2001; {\it Lodato
and Rice}, 2004, 2005).

Radiative Cooling. The published literature on this so far comes from
only three research groups ({\it Nelson et al.}, 2000; {\it Boss}, 2001,
2002b, 2004a; {\it Mej\'ia}, 2004; {\it Cai et al.}, 2006), but work by others
is in progress. Because Solar System-sized disks
encompass significant volumes with small and large optical depth,
this becomes a difficult 3D radiative hydrodynamics
problem. Techniques will be discussed in Section 3.2. For a disk
spanning the conventional planet-forming region, the opacity is due
primarily to dust. Complications which have to be considered include
the dust size distribution, its composition, grain growth and
settling, and the occurrence of fast cooling due to convection. 

In general, the radiative cooling time is dependent on temperature $T$
and metallicity $Z$. Let $\kappa_{r} \sim ZT^{\beta_r}$ and
$\kappa_{p} \sim ZT^{\beta_p}$ be the Rosseland and Planck mean
opacities, respectively, and let $\tau \sim \kappa_r H$ be the
vertical optical depth to the midplane. For large $\tau$,
\begin{equation}\label{eq:tcool-thick}
t_{cool} \sim T/{T_{eff}}^4 \sim T^{-3}\tau \sim T^{-2.5+\beta_r}Z;
\end{equation}
for small $\tau$,
\begin{equation}\label{eq:tcool-thin}
t_{cool} \sim T/\kappa_pT^4 \sim T^{-3-\beta_p}/Z. 
\end{equation}
For most temperatures regimes, we expect $-3 < \beta < 2.5$, so that
$t_{cool}$ increases as $T$ decreases. As $Z$ increases, $t_{cool}$
increases in optically thick regions, but decreases in optically thin
ones.

\noindent
{\it 2.3.2 Heating}

In addition to the internal heating caused by GI's
through shocks, compression, and mass transport, there can be heating
due to turbulent dissipation ({\it Nelson et al.},
2000) and other sources of shocks. In addition, a disk may be
exposed to one or more external radiation fields due to a nearby OB
star (e.g., {\it Johnstone et al.}, 1998), an infalling envelope
(e.g., {\it D'Alessio et al.}, 1997), or the central star (e.g., {\it
Chiang and Goldreich}, 1997).  These forms of heat input can be
comparable to or larger than internal sources of heating and can
influence $Q$ and the surface boundary conditions. Only crude
treatments have been done so far for envelope irradiation ({\it Boss}
2001, 2002b; {\it Cai et al.}, 2006) and for stellar irradiation ({\it
Mej\'ia}, 2004). 

\bigskip
\noindent
\textbf {2.4 Fragmentation}
\bigskip

As shown first by {\it Gammie} (2001) for local thin-disk calculations
and later confirmed by {\it Rice et al.} (2003b) and {\it Mej\'ia et
al.} (2005) in full 3D hydro simulations, disks with a fixed
$t_{cool}$ fragment for sufficiently fast cooling, specifically when
$t_{cool}\Omega \lesssim 3$, or, equivalently, $t_{cool} \lesssim$
$P_{rot}/2$. Finite thickness has a slight stabilizing influence ({\it
Rice et al.}, 2003b; {\it Mayer et al.}, 2004a). When dealing with
realistic radiative cooling, one cannot apply this simple
fragmentation criterion to arbitrary initial disk models. One has to
apply it to the asymptotic phase after nonlinear behavior is
well-developed ({\it Johnson and Gammie}, 2003). Cooling times can be
much longer in the asymptotic state than they are initially ({\it Cai
et al.}, 2006, {\it Mej\'ia et al.}, in preparation). For disks evolved
under isothermal conditions, where a simple cooling time cannot be
defined, local thin-disk calculations show fragmentation when $Q
\lesssim 1.4$ ({\it Johnson and Gammie}, 2003). This is roughly
consistent with results from global simulations (e.g., {\it Boss},
2000; {\it Nelson et al.}, 1998; {\it Pickett et al.}, 2000a, 2003;
{\it Mayer et al.}, 2002, 2004a). Fig. \ref{clump} shows a classic 
example of a fragmenting disk.

\begin{figure}[ht]
\centerline{\epsfig{figure=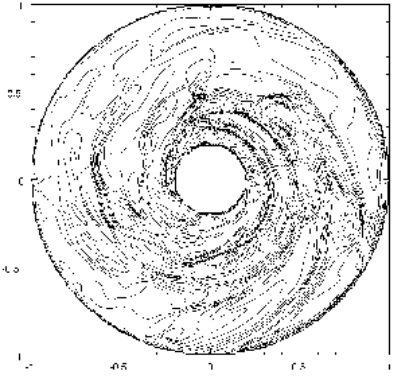,width=60.0mm}}
\caption{\small  Midplane density contours for the isothermal 
evolution of a 0.09 $M_{\odot}$ disk around a 1 $M_{\odot}$ star. 
A multi-Jupiter mass clump forms near 12 o'clock by 374 years. 
The frame in the figure is 40 AU on a side. The figure is adapted from 
{\it Boss} (2000).} 
\label{clump}
\end{figure}

Although there is agreement on conditions for fragmentation, two
important questions remain. Do real disks ever cool fast enough for
fragmentation to occur, and do the fragments last long enough to
contract into permanent protoplanets before being disrupted by tidal
stresses, shear stresses, physical collisions, and shocks?

\section{\textbf {NUMERICAL METHODS}}

A full understanding of disk evolution and the planet formation
process cannot easily be obtained using a purely analytic approach. 
Although numerical methods are powerful, they have flaws and limitations 
that must be taken into account when interpreting results. Here we 
describe some commonly used numerical techniques and their 
limitations.

\bigskip
\noindent
\textbf {3.1 Hydrodynamics}
\bigskip

Numerical models have been implemented using one or the other 
of two broad classes of techniques to solve the hydrodynamic 
equations. Each class discretizes the system in
fundamentally different ways. On one hand, there are particle-based
simulations using Smoothed Particle Hydrodynamics (SPH) ({\it Benz}, 
1990; {\it Monaghan}, 1992), and, on the other, grid-based techniques 
(e.g., {\it Tohline}, 1980; {\it Fryxell et al.}, 1991; {\it Stone and Norman}, 
1992; {\it Boss and Myhill}, 1992; {\it Pickett}, 1995). 

SPH uses a collection of particles distributed in space to represent
the fluid. Each particle is free to move in response to forces
acting on it, so that the particle distribution changes with the
system as it evolves. The particles are collisionless, meaning
that they do not represent actual physical entities, but rather
points at which the underlying distributions of mass, momentum, and 
energy are sampled. In order to calculate hydrodynamic quantities
such as mass density or pressure forces, contributions from other
particles within a specified distance, the {\sl smoothing length}, are
weighted according to a {\sl smoothing kernel} and summed in pairwise
fashion. Mutual gravitational forces are calculated by organizing
particles into a tree, where close particles are treated more
accurately than aggregates on distant branches.

Grid-based methods use a grid of points, usually fixed in
space, on which fluid quantities are defined. In the class of
{\sl finite difference} schemes, fluxes of mass, momentum, and energy
between adjacent cells are calculated by taking finite differences of
the fluid quantities in space. Although not commonly used in
simulations of GI's, the Piecewise Parabolic Method (PPM) of {\it Collela
and Woodward} (1984) represents an example of the class of {\sl finite
volume} schemes. For our purposes, an
important distinguishing factor is that while finite difference and SPH 
methods may require artificial viscosity terms to be added to the equations 
to ensure numerical stability and produce correct dissipation in shocks, 
PPM does not.

\bigskip
\noindent
\textbf {3.2 Radiative Physics}
\bigskip

In Section 2.3, we describe a number of processes by which 
disks may heat and cool. In this section, we discuss various code 
implementations and their limitations.

Fixed EOS evolution is computationally efficient because it removes 
the need to solve an equation for the energy balance. On the
other hand, the gas instantly radiates away all heating due to shocks
and, for the isothermal case, due to compressional heating as well. As
a consequence, the gas may compress to much higher densities than 
are realistic, biasing a simulation towards GI growth and
fragmentation even when a physically appropriate temperature or
entropy scale is used. Although fixed $t_{cool}$'s represent a clear
advance over fixed EOS's, equations \ref{eq:tcool-thick} and
\ref{eq:tcool-thin} show that increasing the temperature, which makes 
the disk more stable, also decreases $t_{cool}$. So it is incorrect to 
view short global cooling times as necessarily equivalent to more rapid 
GI growth and fragmentation. In order for fragmentation to occur, one 
needs {\sl both} a short $t_{cool}$ {\sl and} a disk that is cool enough 
to be unstable (e.g., {\it Rafikov}, 2005). 

The most physically inclusive simulations to date employ
radiative transport schemes that allow $t_{cool}$ to be determined by
disk opacity. Current implementations (Section 2.3) employ variants
of a radiative diffusion approximation in regions of medium to high
optical depth $\tau$, integrated from infinity toward the disk midplane. 
On the other hand, radiative losses actually occur from regions 
where $\tau \lesssim 1$, and so the treatment of the
interface between optically thick and thin regions strongly
influences cooling. Three groups have implemented different
approaches.

{\it Nelson et al.} (2000) assume that the vertical structure of the disk
can be defined at each point as an atmosphere in thermal
equilibrium. In this limit, the interface can be defined by the location 
of the disk {\sl photosphere}, where $\tau=2/3$ (see,
e.g., {\it Mihalas}, 1977). Cooling at each point is then defined as that due
to a blackbody with the temperature of the photosphere. {\it Boss}
(2001, 2002b, 2004a, 2005) performs a 3D flux-limited radiative diffusion 
treatment for the optically thick disk interior ({\it Bodenheimer et al.}, 
1990), coupled to an outer boundary condition where the temperature is set to 
a constant for $\tau < 10$, $\tau$ being measured along the radial direction. 
{\it Mej\'ia} (2004) and {\it Cai et al.} (2006) use the same radiative 
diffusion treatment as {\it Boss} in their disk interior, but they define 
the interface using $\tau = 2/3$, measured vertically,
above which an optically thin atmosphere
model is self-consistently grafted onto the outward flux from the
interior. As discussed in Section 4.2, results for
the three groups differ markedly, indicating that better
understanding of radiative cooling at the disk surface will be
required to determine the fate of GI's. 

\bigskip
\noindent
\textbf{3.3 Numerical Issues}
\bigskip

The most important limitations facing numerical simulations are
finite computational resources. Simulations have a limited duration 
with a finite number of particles or cells, and they must have 
boundary conditions to describe behavior outside the region being
computed. A simulation must distribute grid cells or particles over 
the interesting parts of the system to resolve the relevant physics 
and avoid errors associated with incorrect treatment of the
boundaries. Here we describe a number of requirements for valid
simulations and pitfalls to be avoided.

For growth of GI's, simulations must be able to resolve the wavelengths
of the instabilities underlying the fragmentation. {\it Bate and Burkert}
(1997) and {\it Truelove et al.} (1997) each define criteria based on the
collapse of a Jeans unstable cloud that links a minimum number of grid
zones or particles to either the physical wavelength or mass
associated with Jeans collapse. {\it Nelson} (2006) notes that a Jeans
analysis may be less relevant for disk systems because they are
flattened and rotating rather than homogeneous and instead proposes a
criterion based on the Toomre wavelength in disks. Generally, grid-based 
simulations must resolve the appropriate local
instability wavelength with a minimum of 4 to 5 grid zones in each
direction, while SPH simulations must resolve the local Jeans or 
Toomre mass with a minimum of a few hundred particles. 

Resolution of instability wavelengths will be insufficient to ensure
validity if either the hydrodynamics or gravitational forces are in
error. For example, errors in the hydrodynamics may develop in 
SPH and finite difference methods because a viscous heating term 
must be added artificially to model shock dissipation and, in some
cases, to ensure numerical stability. In practice, the magnitude of 
dissipation depends in part on cell dimensions rather than just on 
physical properties. Discontinuities may be smeared over as many as 
$\sim 10$ or more cells, depending on the method. Further, {\it Mayer et al.} 
(2004a) have argued that because it takes the form of an additional 
pressure, artificial viscosity may by itself reduce or eliminate 
fragmentation. On the other hand, artificial viscosity can promote the 
longevity of clumps (see Fig. 3 of {\it Durisen}, 2006).

Gravitational force errors develop in grid simulations from at
least two sources. First, when {\it Pickett et al.} (2003) place a
small blob within their grid, errors occur in the self-gravitation 
force of the blob that depend on whether the cells containing it 
have the same spacing in each coordinate dimension. Ideally, 
grid zones would have comparable spacing in all directions, but 
disks are both thin and radially extended. Use of spherical and
cylindrical grids tends to introduce disparity in grid spacing. 
Second, {\it Boss} (2000) shows that maximum densities inside clumps 
are enhanced by orders of magnitude as additional terms in his 
Poisson solver, based on a $Y_{lm}$ decomposition, are included. 
SPH simulations encounter a different source of error because
gravitational forces must be softened in order to preserve the
collisionless nature of the particles. {\it Bate and Burkert} (1997) and
{\it Nelson} (2006) each show that large imbalances between the
gravitational and pressure forces can develop if the length scales 
for each are not identical, possibly inducing fragmentation in
simulations. On the other hand, spatially and temporally variable
softening implies a violation of energy conservation. Quantifying
errors from sources such as insufficiently resolved shock dissipation
or gravitational forces cannot be reliably addressed except by
experimentation. Results of otherwise identical simulations performed at
several resolutions must be compared, and identical models must be 
realized with more than one numerical method (as in Section 4.4), so 
that deficiencies in one method can be checked against strengths in 
another.

The disks relevant for GI growth extend over several orders of
magnitude in radial range, while GI's may develop large amplitudes only
over some fraction of that range. Computationally affordable
simulations therefore require both inner and outer radial boundaries,
even though the disk may spread radially and spiral waves propagate up
to or beyond those boundaries. In grid-based simulations, {\it Pickett et
al.} (2000b) demonstrate that numerically induced fragmentation can
occur with incorrect treatment of the boundary. Studies of disk evolution 
must ensure that treatment of the boundaries 
does not produce artificial effects. 

In particle simulations, where there is no requirement that a grid be
fixed at the beginning of the simulation, boundaries are no less a
problem. The smoothing in SPH requires that the distribution of
neighbors over which the smoothing occurs be relatively evenly
distributed in a sphere around each particle for the hydrodynamic
quantities to be well defined. At currently affordable resolutions
($\sim10^5-10^6$ particles), the smoothing kernel extends over a 
large fraction of a disk scale height, so meeting this requirement is
especially challenging. Impact on the outcomes of simulations has 
not yet been quantified. 

\begin{figure*}
\epsscale{1.8}
\plotone{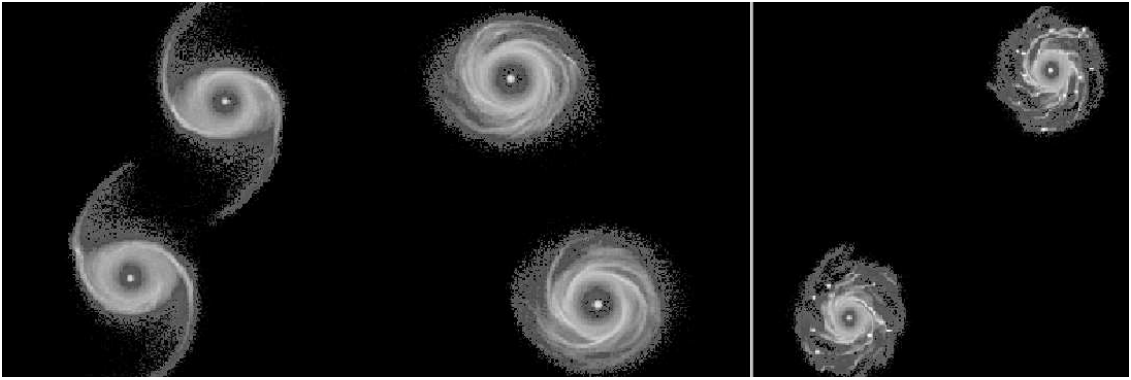}
\caption{\small 
Face-on density maps for two simulations of interacting 
$M=0.1 M_{\odot}$ protoplanetary disks in binaries with 
$t_{cool} = 0.5P_{rot}$ viewed face-on. The binary in the left panel 
has a nearly circular binary orbit with an initial separation of 60 AU 
and is shown after first pericentric passage at 150 yrs (left) 
and then at 450 yrs (right). Large tidally induced spiral arms are 
visible at 150 yrs. The right panel shows a snapshot at 160 yrs from 
a simulation starting from an initial orbital separation that is twice 
as large. In this case, fragmentation into permanent clumps occurs after 
a few disk orbital times. Figures adapted from {\it Mayer et al.} (2005).}
\label{bin}
\end{figure*}

\section{\textbf{KEY ISSUES}}

\noindent
\textbf{4.1 Triggers for GI's}
\bigskip

When disks become unstable, they may either fragment or enter 
a self-regulated phase depending on the cooling time. It is 
therefore important to know how and when GI's may arise in real disks 
and the physical state of the disk at that time. Various mechanisms for 
triggering GI's are conceivable, but only a few have yet been studied 
in any detail. Possibilities include -- the formation of a massive disk 
from the collapse of a protostellar cloud (e.g., {\it Laughlin and 
Bodenheimer}, 1994; {\it Yorke and Bodenheimer}, 1999), clumpy infall 
onto a disk ({\it Boss}, 1997, 1998a), cooling of a disk from a stable 
to an unstable state, slow accretion of mass, accumulation 
of mass in a magnetically dead zone, perturbations by a binary companion, 
and close encounters with other star/disk systems ({\it Boffin et al.}, 
1998; {\it Lin et al.}, 1998). A few of these will be discussed further, 
with an emphasis on some new results on effects of binarity.

Several authors start their disks with stable or marginally
stable $Q$-values and evolve them to instability either by slow idealized 
cooling (e.g., {\it Gammie}, 2001; {\it Pickett et al.}, 2003; 
{\it Mej\'ia et al.}, 2005) or by more realistic radiative cooling (e.g., 
{\it Johnson and Gammie}, 2003; {\it Boss}, 2005, 2006; {\it Cai et al.}, 2006). 
To the extent tested, fragmentation in idealized cooling cases are consistent 
with the Gammie criterion (Section 2.4). 
With radiative cooling, as first pointed out by {\it Johnson and Gammie} (2003),
it is difficult to judge whether a disk will fragment when it reaches 
instability based on its initial $t_{cool}$. When {\it Mayer et al.}  
(2004a) grow the mass of a disk while keeping its temperature constant,
dense clumps form in a manner similar to clump formation
starting from an unstable disk. A similar treatment of accretion needs to be 
done using realistic radiative cooling. Simulations like these suggest that,
in the absence of a strong additional source of heating, 
GI's are unavoidable in protoplanetary disks with sufficient 
mass ($\sim 0.1 M_\odot$ for a $\sim 1 M_\odot$ star).

A disk evolving primarily due to magnetorotational instabilites (MRI's) may 
produce rings of cool gas in the disk midplane where the ionization fraction 
drops sufficiently to quell MRI's ({\it Gammie}, 1996; {\it Fleming and Stone}, 
2003). Dense rings associated with these magnetically {\sl dead zones} should 
become gravitationally unstable and may well trigger a localized onset of GI's. 
This process might lead to disk outbursts related to FU Orionis events 
({\it Armitage et al.}, 2001) and induce chondrule-forming 
episodes ({\it Boley et al.}, 2005).

A phase of GI's robust enough to lead to gas giant protoplanet 
formation might be achieved through external triggers, like a binary star 
companion or a close encounter with another protostar and its disk. A few 
studies have explored the effects of binary companions on GI's.
{\it Nelson} (2000) follows the evolution of disks in an equal-mass binary 
system with a semimajor axis of 50 AU and an eccentricity of 0.3 and
finds that the disks are heated by internal shocks and viscous processes 
to such an extent as to become too hot for gas giant planet formation 
either by disk GI's or by core accretion, because volatile ices and organics 
are vaporized. In a comparison of the radiated emission calculated from his
simulation to those from the L1551 IRS5 system, {\it Nelson} (2000) finds 
that the simulation is well below the observed system and therefore that the 
temperatures in the simulation are underestimates. He therefore 
concludes that ``planet formation is unlikely in equal-mass binary systems 
with $a \sim$ 50 AU.'' Currently, over two dozen binary or triple star 
systems have known extrasolar planets, with binary separations
ranging from $\sim 10$ AU to $\sim 10^3$ AU, so some means must
be found for giant planet formation in binary star systems with
relatively small semimajor axes.

Using idealized cooling, {\it Mayer et al.} (2005) find that the effect of 
binary companions depends on the mass of the disks involved and on 
the disk cooling rate. For a pair of massive disks ($M \sim 0.1 M_{\odot}$), 
formation of permanent clumps can be suppressed as a result of intense 
heating from spiral shocks excited by the tidal perturbation (Fig. \ref{bin}
left panel). Clumps do not form in such disks for binary orbits having a 
semimajor axis of $\sim 60$ AU even when $t_{cool} < P_{rot}$. The 
temperatures reached in these disks are $> 200$ K and would vaporize 
water ice, hampering core accretion, as argued by {\it Nelson} (2000). On 
the other hand, pairs of less massive disks ($M \sim 0.05 M_{\odot}$) that 
would not fragment in isolation since they start with $Q \sim 2$, can 
produce permanent clumps provided that $t_{cool} \lesssim P_{rot}$. 
This is because the tidal perturbation is weaker
in this case (each perturber is less massive) and the resulting shock
heating is thus diminished. Finally, the behavior of such binary systems
approaches that seen in simulations of isolated disks once the semimajor
axis grows beyond $100$ AU (Fig. \ref{bin} right panel).

Calculations by {\it Boss} (2006) of the evolution of initially
marginally gravitationally stable disks show that the presence of
a binary star companion could help to trigger the formation of
dense clumps. The most likely explanation for the difference in outcomes 
between the models of {\it Nelson} (2000) and {\it Boss} (2006) is 
the relatively short cooling times in the latter models
($\sim 1$ to 2 $P_{rot}$, see {\it Boss} 2004a) compared to the 
effective cooling time in {\it Nelson} (2000) of $\sim 15P_{rot}$
at 5 AU, dropping to $\sim P_{rot}$ at 15 AU.
Similarly, some differences in outcomes between the results
of {\it Boss} (2006) and {\it Mayer et al.} (2005) can be expected 
based on different choices of the binary semimajor axes and 
eccentricities and differences in the thermodynamics. For example,
{\it Mayer et al.} (2005) turn off cooling in regions with densities
higher than $10^{-10}$ g cm$^{-3}$ to account for high optical 
depths.

Overall, the three different calculations agree that excitation 
or suppression of fragmentation by a binary companion depends 
sensitively on the balance between compressional/shock heating 
and cooling. This balance appears to depend 
on the mass of the disks involved. Interestingly, lighter disks 
are more likely to fragment in binary systems according to both 
{\it Mayer et al.} (2005) and {\it Boss} (2006).

\bigskip
\noindent
\textbf{4.2 Disk Thermodynamics}
\bigskip

As discussed in Sections 2.2 and 4.1, heating and cooling are 
perhaps the most important processes affecting the growth and 
fate of GI's. Thermal regulation in the nonlinear regime leads
naturally to systems near their stability limit where temporary
imbalances in one heating or cooling term lead to a proportionate
increase in a balancing term. For fragmentation to occur, a disk must
cool quickly enough, or fail to be heated for long enough, to upset
this self-regulation. A complete model of the energy balance that includes 
all relevant processes in a time-dependent manner is beyond the capabilities 
of the current generation of models. It requires knowledge of all the 
following -- external radiation sources and their influence on the disk 
at each location, the energy lose rate of the disk due to
radiative cooling, dynamical processes that generate thermal energy
through viscosity or shocks, and a detailed equation of state to 
determine how much heating any of those dynamical processes
generate. Recent progress towards understanding disk evolution
has focused on the more limited goals of quantifying the sensitivity
of results to various processes in isolation. 

In a thin, steady state $\alpha$-disk, the heating and cooling times 
are the same and take a value ({\it Pringle}, 1981; {\it Gammie}, 2001):
\begin{equation}\label{eq:tcool-alpha}
t_{cool} = \left.4\over9\right. \left[\gamma(\gamma-1)\alpha\Omega\right]^{-1}.
\end{equation}
For $\alpha\sim10^{-2}$ and 
$\gamma=1.4$, equation \ref{eq:tcool-alpha} gives $\sim12 P_{rot}$.
This is a crude upper limit on the actual time scale required to 
change the disk thermodynamic state. 
External radiative heating from the star and any remaining
circumstellar material can contribute a large fraction of the total
heating ({\it D'Alessio et al.}, 1998; {\it Nelson et al.}, 2000), as will any
internal heating due to globally generated dynamical instabilities
that produce shocks. Each of these processes actually makes the disk
more stable by heating it, but, as a consequence, dynamical evolution
slows until the disk gains enough mass to become unstable again. The
marginally stable state will then be precariously held because the
higher temperatures mean that all of the heating and cooling
time scales, i.e., the times required to remove or replace all the disk
thermal energy, are short (equations \ref{eq:tcool-thick} and
\ref{eq:tcool-thin}). When the times are short, any disruption of the 
contribution from a single source may be able to change the 
thermodynamic state drastically within only a few orbits, 
perhaps beyond the point where balance can be restored.

A number of models (Section 2.3) have used fixed EOS evolution 
instead of a full solution of an energy equation to explore disk evolution. 
A fixed EOS is equivalant to specifying the outcomes of all heating and
cooling events that may occur during the evolution, short-circuiting
thermal feedback. If, for example, the temperature or entropy is set
much too high or too low, a simulation may predict either that no GI's
develop in a system, or that they inevitably develop and produce 
fragmentation, respectively. Despite this limitation, fixed EOS's have
been useful to delineate approximate boundaries for regions of marginal 
stability. Since the thermal state is fixed, disk stability (as quantified 
by equation \ref{eq:Toomre-Q}) is essentially determined by the disk's mass 
and spatial dimensions, though its surface density. Marginal stability
occurs generally at $Q \approx$ 1.2 to 1.5 for locally isentropic evolutions, 
with a tendency for higher $Q$'s being required to 
ensure stability with {\sl softer} EOS's (i.e., with lower $\gamma$ values) 
({\it Boss} 1998a; {\it Nelson et al.}, 1998; {\it Pickett et al.}, 1998, 2000a;
{\it Mayer et al.}, 2004a). At temperatures appropriate for observed systems
(e.g., {\it Beckwith et al.}, 1990), these $Q$ values correspond to disks 
more massive than $\sim0.1M_*$ or surface densities
$\Sigma\gtrsim10^3$~gm/cm$^2$.

As with their fixed EOS cousins, models with fixed $t_{cool}$ 
can quantify boundaries at which fragmentation may set in. 
They represent a clear advance over fixed EOS evolution by
allowing thermal energy generated by shocks or compression to be 
retained temporarily, and thereby enabling the disk's natural thermal 
regulation mechanisms to determine the evolution. Models that employ 
fixed cooling times can address the question of how violently the disk's 
thermal regulation mechanisms must be disrupted before they can no 
longer return the system to balance. An example of the value of fixed
$t_{cool}$ calculations is the fragmentation criterion
$t_{cool} \lesssim 3\Omega^{-1}$ (see Section 2.4).

The angular momentum transport associated with disk self-gravity is a
consequence of the gravitational torques induced by GI spirals
(e.g., {\it Larson}, 1984). 
The viscous $\alpha$ parameter is actually a measure of that
stress normalized by the local disk pressure. As shown in equation
\ref{eq:tcool-alpha} and reversing the positions of $t_{cool}$ and
$\alpha$, the stress in a self-gravitating disk depends on the cooling
time and on the equation of state through the specific heat ratio.
As long as the dimensionless scale height is 
$H \lesssim 0.1$, global simulations by {\it Lodato and Rice} (2004) 
with $t_{cool}\Omega =$ constant confirm Gammie's assumption 
that transport due to disk self-gravity can be modeled as a local
phenomenon and that equation \ref{eq:tcool-alpha} is accurate. 
{\it Gammie} (2001) and {\it Rice et al.} (2005) 
show that there is a maximum stress that can be supplied by such a
quasi-steady, self-gravitating disk. Fragmentation occurs if the stress 
required to keep the disk in a quasi-steady state exceeds this maximum 
value. The relationship between the stress and the specific heat ratio, 
$\gamma$, results in the cooling time required for fragmentation increasing 
as $\gamma$ decreases.  For $\gamma = 7/5$, the cooling time below 
which fragmentation occurs may be more like $2 P_{rot}$, not the 
$3/\Omega \approx P_{rot}/2$  obtained for $\gamma = 2$ 
({\it Gammie}, 2001; {\it Mayer et al.}, 2004b; {\it Rice et al.}, 2005).

Important sources of stress and heating in the disk, that lie 
outside the framework of Gammie's local analysis, are global
gravitational torques due to low-order GI spiral modes. There
are two ways this can happen -- a geometrically thick massive
disk ({\it Lodato and Rice}, 2005) and a fixed global $t_{cool} =$
constant ({\it Mej\'ia et al.}, 2005). Disks then initially produce 
large-amplitude spirals, resulting in a transient burst of global mass 
and angular momentum redistribution. For $t_{cool} =$ constant and 
moderate masses, the disks then settle down to a self-regulated 
asymptotic state but with gravitational stresses significantly
higher than predicted by equation \ref{eq:tcool-alpha}
({\it Michael et al.}, in preparation). 
For the very massive $t_{cool}\Omega =$ constant disks, recurrent episodic
redistributions occur. In all these cases, the heating in spiral shocks
is spatially and temporally very inhomogeneous, as are fluctuations
in all thermodynamic variables and the velocity field. 

The most accurate method to determine the internal thermodynamics 
of the disk is to couple the equations of radiative transport to the
hydrodynamics directly. All heating or cooling due to radiation will
then be properly defined by the disk opacity, which depends on local
conditions. This is important because some
fraction of the internal heating will be highly inhomogeneous, occurring
predominantly in compressions and shocks as gas enters a
high density spiral structure, or at high altitudes where waves from
the interior are refracted and steepen into shocks ({\it Pickett et al.}
2000a) and where disks may be irradiated ({\it Mej\'ia}, 2004; 
{\it Cai et al.}, 2006). Temperatures and the $t_{cool}$'s that depend 
on them will then be neither simple functions of radius, nor a single 
globally defined value. Depending on whether the local cooling time 
of the gas inside the high density spiral structure is short enough,
fragmentation will be more or less likely, and additional hydrodynamic
processes such as convection may become active if large enough
gradients can be generated. 

Indeed, recent simulations of {\it Boss} (2002a, 2004a) suggest that 
vertical convection is active in disks when radiative transfer is included, 
as expected for high $\tau$ according to {\it Ruden and Pollack} (1991).
This is important because convection will keep the upper layers of the
disk hot, at the expense of the dense interior, so that radiative
cooling is more efficient and fragmentation is enhanced. 
The results have not yet been confirmed by other work and
therefore remain somewhat controversial. Simulations by {\it Mej\'ia} 
(2004) and {\it Cai et al.} (2006) are most similar to those of
{\it Boss} and could have developed convection sufficient to
induce fragmentation, but none seems to occur. 
No fragmentation occurs in {\it Nelson et al.} (2000)
either, where convection is implicitly assumed to be efficient 
through their assumption that the entropy of each vertical column is 
constant. Recent re-analysis of their results reveals 
$t_{cool}\sim3$ to 10 $P_{rot}$, depending on radius, which 
is too long to allow fragmentation. These $t_{cool}$'s are in agreement 
with those seen by {\it Cai et al.} (2006) and by {\it Mej\'ia et al.} 
(in preparation) for solar metallicity.
The {\it Nelson et al.} results are also interesting because their
comparison of the radiated output to SEDs observed for real
systems demonstrates that substantial additional heating beyond
that supplied by GI's is required to reproduce the observations,
perhaps further inhibiting fragmentation in their models. However,
using the same temperature distribution between 1 and 10 AU now
used in Boss's GI models, combined with temperatures outside 
this region taken from 
models by {\it Adams et al.} (1988), {\it Boss and Yorke} 
(1996) are able to reproduce the SED of the T Tauri system. It is 
unclear at present why their results differ from those of 
{\it Nelson et al.}  (2000).
 
The origins of the differences between the three studies are uncertain,
but possibilities include differences of both numerical and physical
origin. The boundary treatment at the optically thick/thin interface
is different in each case (see Section 3.2), influencing the efficiency of
cooling, as are the numerical methods and resolutions. {\it Boss} and
the {\it Cai/Mej\'ia} group each use 3D grid codes but with 
spherical and cylindrical grids respectively, and each with a different 
distribution of grid zones, while {\it Nelson et al.} use a 2D SPH code.
Perhaps significantly, {\it Cai/Mej\'ia} assume their
ideal gas has $\gamma = 5/3$ while {\it Boss} adopts an EOS that 
includes rotational and vibrational states of hydrogen, so that 
$\gamma \approx 7/5$ for typical disk conditions. It is possible that 
differences in the current results may be explained if the same 
sensitivities to $\gamma$ seen in fixed EOS and fixed cooling
simulations also hold when radiative transfer is included. {\it Boss
and Cai} (in preparation) are now conducting direct comparison calculations to
isolate the cause of their differences. The preliminary indication is that
the radiative boundary conditions may be the critical factor.

Discrepant results for radiatively cooled models should not overshadow 
the qualitative agreement reached about the relationship between disk
thermodynamics and fragmentation. If the marginally unstable state of 
a self-regulated disk is upset quickly enough by an increase in cooling 
or decrease in heating, the disk may fragment. What is still very unclear 
is whether such conditions can develop in real planet-forming disks.
It is key to develop a full 3D portrait of the disk surface, so that
radiative heating and cooling sources may be included self-consistently 
in numerical models. Important heating sources will
include the envelope, the central star, neighboring stars, and 
self-heating from other parts of the disk, all of which will be sensitive 
to shadowing caused by corrugations in the disk surface that develop 
and change with time due to the GI's themselves.
Preliminary studies of 3D disk structure ({\it Boley and Durisen},
2006) demonstrate that vertical distortions, analogous to
hydraulic jumps, will in fact develop (see also {\it Pickett et al.}, 2003). 
If these corrugations are sufficient to cause portions of the
disk to be shadowed, locally rapid cooling may occur in the shadowed
region, perhaps inducing fragmentation.

An implicit assumption of the discussion above is that the
opacity is well known. In fact, it is not. The dominant source of
opacity is dust, whose size distribution, composition, and spatial
distribution will vary with time ({\it Cuzzi et al.}, 2001; {\it Klahr}, 
2003, see also Section 5 below),
causing the opacity to vary as a result. So far, no models of GI
evolution have included effects from any of these processes, except 
that {\it Nelson et al.} model dust destruction while 
{\it Cai} and {\it Mej\'ia} consider opacity due to large grains. Possible
consequences are a misidentification of the disk photospheric surface
if dust grains settle towards the midplane, or incorrect radiative
transfer rates in optically thick regions if the opacities themselves
are in error.

\begin{figure*}
\epsscale{1.6}
\plottwo{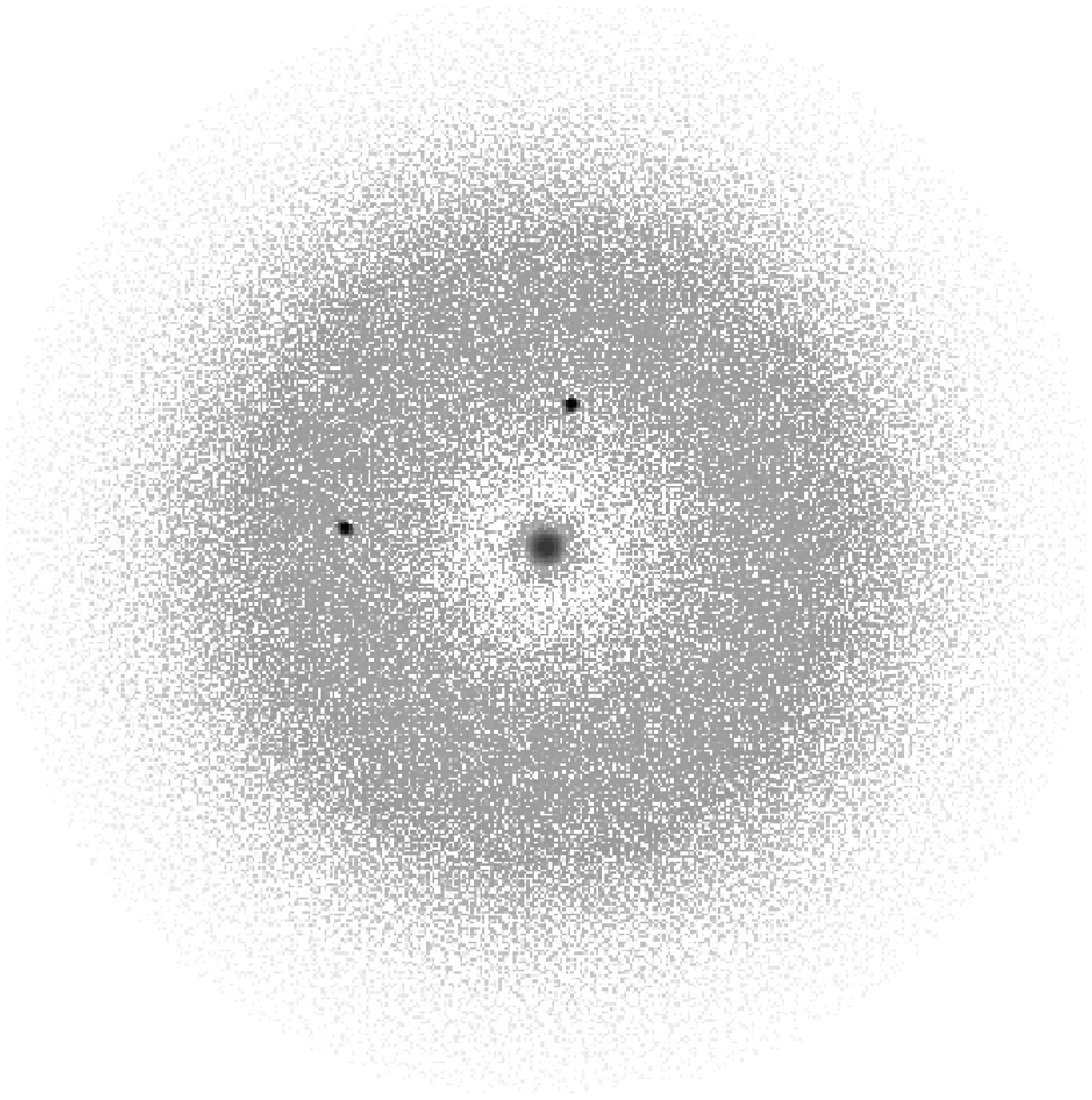}{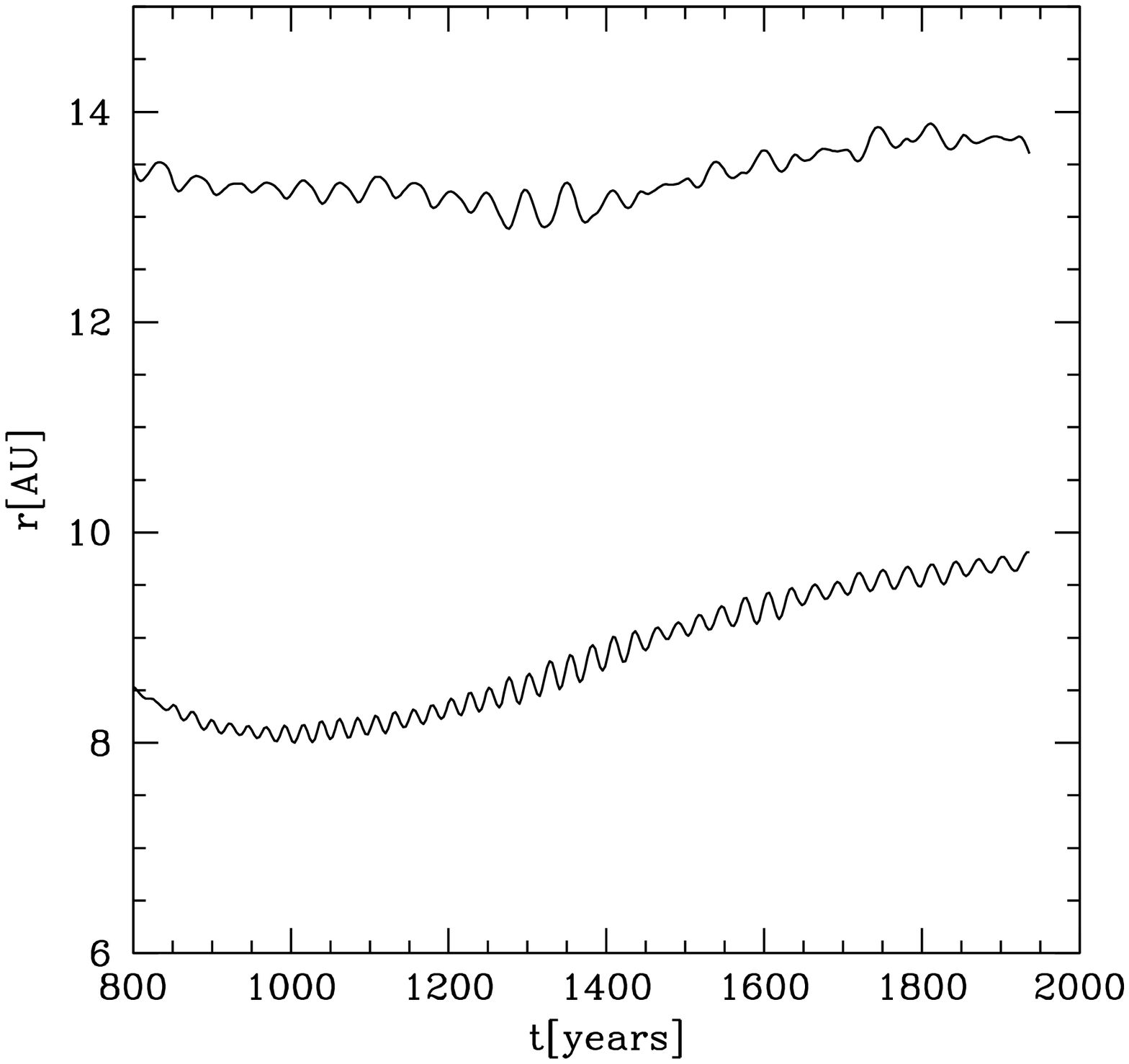}
\caption{\label{fig:lucio-fig}
\small The orbital evolution of two clumps (right) formed in a
massive, growing protoplanetary disk simulation described in Mayer et
al. (2004). A face-on view of the system after 2264 years of evolution
is shown on the left, using a color coded density map (the box is 38
AU on a side). In the right panel, the orbital evolution of the two
clumps is shown. Overal, both clumps migrate outward.}  
\end{figure*}

\bigskip
\noindent
\textbf{4.3 Orbital Survival of Clumps}
\bigskip

Once dense clumps form in a gravitationally unstable disk, the
question becomes one of survival: Are they transient structures or
permanent precursors of giant planets? Long-term 
evolution of simulations that develop clumps is difficult because
it requires careful consideration of not only the large-scale dynamical
processes that dominate formation but also physical processes that
exert small influences over long time scales (e.g., migration and
transport due to viscosity). It also requires that boundary conditions
be handled gracefully in cases where a clump or the disk itself tries
to move outside the original computational volume.

On a more practical level, the extreme computational cost of
performing such calculations limits the time over which systems 
may be simulated. As a dense clump forms, the temperatures, densities, 
and fluid velocities within it all increase. As a result, time steps, limited
by the Courant condition, can decrease to as little as minutes or hours as the 
simulation attempts to resolve the clump's internal structure. So far only 
relatively short integration times of up to a few$\times 10^3$ yrs have 
been possible. Here, we will focus on the results of 
simulations and refer the reader to the chapters by {\it Papaloizou et al.} 
and {\it Levison et al.} for discussions of longer-term interactions. 

In the simplest picture of protoplanet formation via GI's, structures
are assumed to evolve along a continuum of states
that are progressively more susceptible to fragmentation, 
presumably ending in one or more bound objects which eventually 
become protoplanets. {\it Pickett et al.} (1998, 2000a, 2003) and 
{\it Mej\'ia et al.} (2005) simulate initially smooth disks subject to 
growth of instabilities and, indeed, find growth of large-amplitude 
spiral structures that later fragment into arclets or clumps. Instead of 
growing more and more bound, however, these dense structures 
are sheared apart by the background flow within an orbit or less, 
especially when shock heating is included via an
artificial viscosity. This suggests that a detailed understanding of
the thermodynamics inside and outside the fragments is critical
for understanding whether fragmentation results in permanently 
bound objects.

Assuming that permanently bound objects do form, two additional
questions emerge. First, how do they accrete mass and how much
do they accrete? Second, how are they influenced by the
remaining disk material?  Recently, {\it Mayer et al.} (2002, 2004a) and
{\it Lufkin et al.} (2004) have used SPH calculations to follow the
formation and evolution of clumps in simulations covering up to  
50 orbits (roughly 600 yrs), and {\it Mayer et al.} (in preparation) 
are extending these calculations to several thousand years. They find 
that, when a locally isothermal EOS is used well past initial 
fragmentation, clumps grow to $\sim10M_J$ within a few hundred 
years. On the other hand, in simulations using an ideal gas 
EOS plus bulk viscosity, accretion rates are much lower 
($<10^{-6} M_{\odot}/$yr), and clumps do not grow to more than a few
$M_J$ or $\sim1$\% of the disk mass. The assumed
thermodynamic treatment has important effects not only on the 
survival of clumps, but also on their growth.

{\it Nelson and Benz} (2003), using a grid-based code and
starting from a 0.3$M_J$ seed planet, show that accretion rates this
fast are unphysically high because the newly accreted gas
cannot cool fast enough, even with the help of convection, unless some
localized dynamical instability is present in the clump's envelope.
So, the growth rate of an initially small protoplanet may be limited
by its ability to accept additional matter rather than the disk's
ability to supply it. They note (see also {\it Lin and Papaloizou}, 1993;
{\it Bryden et al.}, 1999; {\it Kley}, 1999; {\it Lubow et al.}, 1999; 
{\it Nelson et al.}, 2000) that the accretion process after formation is 
self-limiting at a mass comparable to the largest planet masses yet 
discovered (see the chapter by {\it Udry et al.}). 

Fig. \ref{fig:lucio-fig} shows one of the extended {\it Mayer et al.}
simulations, containing two clumps in one disk realized with $2 \times
10^5$ particles, and run for about 5,000 years (almost 200 orbits at
$10$ AU). There is little hint of inward orbital migration over a few
thousand year time scale. Instead, both clumps appear to migrate slowly
outward. {\it Boss} (2005) uses sink particles (``virtual planets'') 
to follow a clumpy disk for about 1,000 years. He
also finds that the clumps do not migrate rapidly. In both works, 
the total simulation times are quite short compared to the
disk lifetime and so are only suggestive of the longer-term fate of
the objects. Nevertheless, the results are important, because they
illustrate shortcomings in current analytic models of migration.

Although migration theory is now extremely well developed 
(see the chapter by {\it Papaloizou et al.}), predictions for migration
at the earliest phases of protoplanet formation by GI's are
difficult to make, because many of the assumptions 
on which the theory is based
are not well satisfied. More than one protoplanet may form in the same
disk, they may form with masses larger than linear theory can
accommodate, and they may be significantly extended rather than the
point masses assumed by theory. If the disk remains massive, it may
also undergo gravitoturbulence that changes the disk's mass 
distribution on a short enough time scale to call into question the 
resonance approximations in the theory. If applicable in
the context of these limitations, recent investigations into the
character of corotation resonances (see the chapter by {\it Papaloizou et al.})
and vortex excitation ({\it Koller et al.}, 2003) in the corotation region
may be of particular interest, because a
natural consequence of these processes is significant mass
transport across the clump's orbit and reduced inward migration, 
which is in fact seen in the above simulations.

\begin{figure*}
\epsscale{1.25}
\plotone{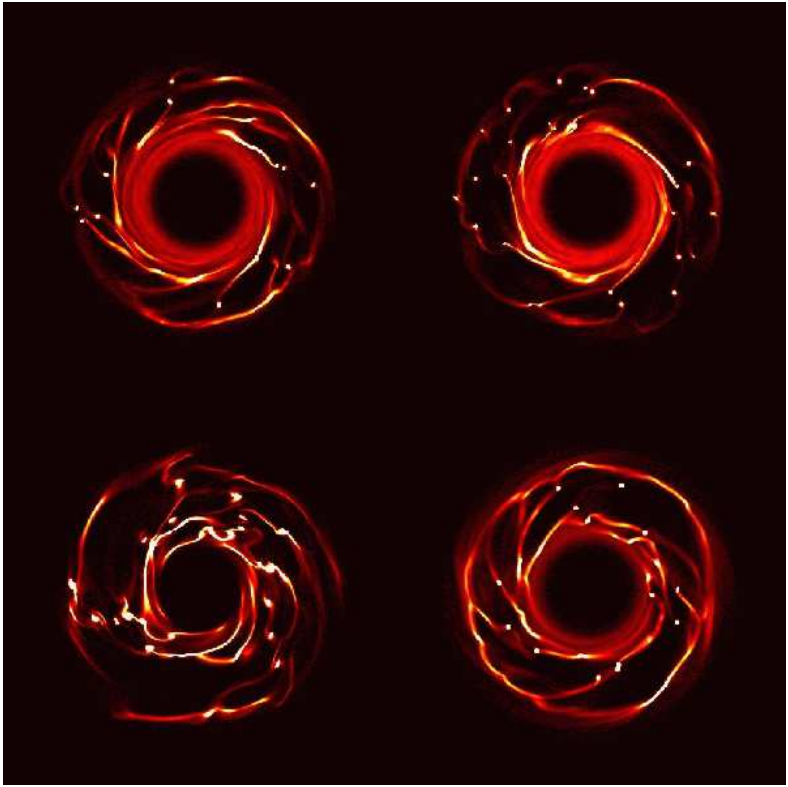}
\caption{\small Equatorial slice density maps of the disk in the test runs after 
about 100 yrs of evolution. The initial disk is 20 AU in diameter. From
top left to bottom right are the results from GASOLINE and GADGET2 (both SPH codes), from the Indiana cylindrical-grid code, and from the AMR Cartesian-grid code 
FLASH. The SPH codes adopt the shear-reduced artificial viscosity of
{\it Balsara} (1995).} 
\label{comp}
\end{figure*}

\bigskip
\noindent
\textbf{4.4 Comparison Test Cases}
\bigskip

Disk instability has been studied so far with various types
of grid codes and SPH codes that have different relative strengths 
and weaknesses (Section 3). Whether different numerical techniques 
find comparable results with nearly identical assumptions is not yet known,
although some comparative studies have been attempted ({\it Nelson
et al.}, 1998). Several aspects of GI behavior can be highly
dependent on code type. For example, SPH codes require artificial viscosity 
to handle shocks such as those occurring along spiral arms. 
Numerical viscosity can smooth out the velocity field in overdense regions, 
possibly inhibiting collapse ({\it Mayer et al.}, 2004a) but, at the same time,
possibly increasing clump longevity if clumps form (see Fig. 3 of
{\it Durisen}, 2006). Gravity solvers that are both accurate and fast  
are a robust feature of SPH codes, while gravity solvers in grid codes can 
under-resolve the local self-gravity of the gas ({\it Pickett et al.}, 2003). Both
types of codes can lead to spurious fragmentation or suppress it when a
force imbalance between pressure and gravity results at scales 
comparable to the local Jeans or Toomre length due to lack of resolution 
({\it Truelove et al.}, 1997; {\it Bate and Burkert}, 1997; {\it Nelson}, 2006).  

Another major code difference is in the set up of initial conditions. Although
both Eulerian grid-based and Lagrangian particle-based techniques represent 
an approximation to the continuum fluid limit, noise levels 
due to discreteness are typically higher in SPH simulations. 
Initial perturbations are often applied in grid-based simulations to
seed GI's (either random or specific modes or both, e.g., {\it Boss}, 1998a), but
are not required in SPH simulations, because they already have built-in
Poissonian noise at the level of $\sqrt{N}/N$ or more, where $N$ is
the number of particles. In addition, the SPH calculation of hydrodynamic
variables introduces small scale noise at the level of $1/N_{neigh}$,
where $N_{neigh}$ is the number of neighboring particles contained in
one smoothing kernel. Grid-based simulations require
boundary conditions which restrict the dynamic range of the simulations.
For example, clumps may reach the edge of a computational volume after 
only a limited number of orbits ({\it Boss}, 1998a, 2000; {\it Pickett et al.}, 2000a). 
Cartesian grids can lead to artificial diffusion of angular momentum in a disk,
a problem that can be avoided using a cylindrical grid
({\it  Pickett et al.}, 2000a) or spherical grid ({\it Boss and Myhill}, 1992). 
{\it Myhill \& Boss} (1993) find good agreement between 
spherical and Cartesian grid results for a nonisothermal rotating protostellar 
collapse problem, but evolution of a nearly equilibrium disk over many orbits 
in a Cartesian grid is probably still a challenge.

In order to understand how well different numerical techniques can agree
on the outcome of GI's, different codes need to run the same initial 
conditions. This is being done in a large, on-going code-comparison project that
involves eight different codes, both grid-based and SPH. Among the grid codes, 
there are several adaptive mesh refinement (AMR) schemes.
The comparison is part of a larger effort involving several areas
of computational astrophysics 
(http://krone.physik.unizh.ch/$\sim$moore/wengen/tests.html). 
The system chosen for the comparison is a uniform temperature, massive,
and initially very unstable disk
with a diameter of about 20 AU.  The disk is evolved isothermally
and has a $Q$ profile that decreases outward,
reaching a minimum value $\sim 1$ at
the disk edge. The disk model is created using
a particle representation by letting its mass grow slowly, as described in
{\it Mayer et al.} (2004a). This distribution is then interpolated onto 
the various grids. 

Here we present the preliminary results of the code comparisons from 
four codes -- two SPH codes called GASOLINE ({\it Wadsley et al.}, 
2004) and GADGET2 ({\it Springel et al.}, 2001; {\it Springel}, 2005), the 
Indiana University code with a fixed cylindrical grid ({\it Pickett}, 1995;
{\it Mej\'ia}, 2004), and the Cartesian AMR code called 
FLASH ({\it Fryxell et al.}, 2000). Readers 
should consult the published literature for detailed descriptions, but we briefly 
enumerate some basic features.
FLASH uses a PPM-based Riemann solver on a Cartesian grid with directional
splitting to solve the Euler equations, and it uses an iterative multi-grid
Poisson solver for gravity. Both GASOLINE and GADGET2 solve the Euler 
equations using SPH and solve gravity using a treecode, a binary tree in the 
case of GASOLINE and an oct-tree in the case of GADGET2. Gravitational forces
from individual particles are smoothed using a spline kernel softening, and 
they both adopt the {\it Balsara} (1995) artificial viscosity that minimizes 
shear forces on large scales. The Indiana code is a finite difference grid-based 
code which solves the equations of hydrodynamics using the Van 
Leer method. Poisson's equation is solved at the end of each hydrodynamic 
step by a Fourier transform of the density in the azimuthal direction, direct solution 
by cyclic reduction of the transform in ($r$,$z$), and a transform back to real space
({\it Tohline}, 1980). The code's Von Neumann-Richtmeyer artificial bulk viscosity 
is not used for isothermal evolutions.

The two SPH codes are run with fixed 
gravitational softening, and the local Jeans length (see {\it Bate and Burkert}, 
1997) before and after clump formation is well resolved. 
Runs with adaptive gravitational softening will soon be included
in the comparison. Here we show the results of the runs whose initial
conditions were generated from the  $8 \times 10^5$ particles setup,
which was mapped onto a 512x512x52 Cartesian grid 
for FLASH and onto a 512x1024x64 ($r$,$\phi$,$z$) cylindrical grid for the 
Indiana code. Comparable resolution (cells for grids 
or gravity softening for SPH runs) is available initially in the outer 
part of the disk, where the Q parameter reaches its minimum.  
In the GASOLINE and GADGET2 runs, the maximum spatial resolution is
set by the gravitational softening at 0.12 AU. Below this scale, gravity is
essentially suppressed. The FLASH run has a initial resolution of 
0.12 AU at 10 AU, comparable with the SPH runs.
The Indiana code has the same resolution as FLASH in the radial 
direction but has a higher azimuthal resolution of 0.06 AU at 10 AU.

As it can be seen from Fig. \ref{comp}, the level of agreement between 
the runs is satisfactory, although significant differences are noticeable.
More clumps are seen in the Indiana code simulation. On the other end, 
clumps have similar densities in FLASH and GASOLINE, while they appear 
more fluffy in the Indiana code than in the other three. The causes are 
probably different gravity solvers and the non-adaptive nature of the
Indiana code. Even within a single category of code, SPH
or grid-based,  different types of viscosity, both artificial and numerical, 
might be more or less diffusive and affect the formation and survival of clumps.
In fact, tests show that more fragments are present in SPH runs 
with shear-reduced artificial viscosity than with full shear viscosity. 

Although still in an early stage, the code comparison has 
already produced one important result, namely that, once favorable 
conditions exist, widespread fragmentation is 
obtained in high-resolution simulations using any of the standard 
numerical techniques. On the other hand, 
the differences already noticed require further understanding and will be 
addressed in a forthcoming paper ({\it Mayer et al.}, in preparation).
Although researchers now agree on conditions for disk fragmentation, 
no consensus yet exists about whether or where real disks fragment 
or how long fragments really persist. Answers 
to these questions require advances over current techniques for treating
radiative physics and compact structures in global simulations.

\section{\textbf{INTERACTIONS WITH SOLIDS}}

The standard model for the formation of giant gaseous planets
involves the initial growth of a rocky core that, when
sufficiently massive, accretes a gaseous envelope ({\it Bodenheimer
and Pollack}, 1986; {\it Pollack et al.}, 1996). In this scenario, the
solid particles in the disk must first grow from micron-sized
dust grains to kilometer-sized planetesimals that then
coagulate to form the rocky core.
                                                                                
In a standard protoplanetary disk, the gas pressure near the 
disk midplane will generally decrease with increasing radius resulting
in an outward pressure gradient that causes
the gas to orbit with sub-Keplerian velocities. The solid particles,
on the other hand, do not feel the gas pressure and orbit with 
Keplerian velocities. This velocity difference results in a drag force
that generally causes the solid particles to lose angular momentum
and to spiral inward toward the central star with a radial drift velocity
that depends on the particle size ({\it Weidenschilling}, 1977).

While this differential radial drift can mix together particles of different 
size and allow large grains to grow by sweeping up smaller grains
({\it Weidenschilling and Cuzzi}, 1993), it also introduces a potential
problem. Depending on the actual disk properties, 
the inward radial velocity for particles with sizes
between $1$ cm and $1$ m can be as high as $10^4$ cm s$^{-1}$ 
({\it Weidenschilling}, 1977), so that these particles could easily 
migrate into the central star before becoming large enough to 
decouple from the disk gas. If these particles do indeed have 
short residence times in the disk, it is difficult to envisage how
they can grow to form the larger kilometer-sized planetesimals which
are required for the subsequent formation of the planetary cores.

The above situation is only strictly valid in smooth, laminar disks
with gas pressures that decrease monotonically with increasing
radius.  If there are any regions in the disk that have local
pressure enhancements, the situation can be very different.  
In the vicinity of a pressure enhancement, the gas velocity can 
be either super- and sub-Keplerian depending on the local gas pressure
gradient. The drag force can then cause solid particles to 
drift outwards or inwards, respectively
({\it Haghighipour and Boss}, 2003a,b). The net effect is that the solid
particles should drift towards pressure maxima.
A related idea is that a baroclinic instability could
lead to the production of long-lived, coherent vortices
({\it Klahr and Bodenheimer}, 2003) and 
that solid particles would drift towards the center of the vortex where the
enhanced concentration could lead to accelerated grain growth 
({\it Klahr and Henning}, 1997). The existence of such vortices is, 
however, uncertain ({\it Johnson and Gammie}, 2006).

\begin{figure}[ht]
\centerline{\epsfig{figure=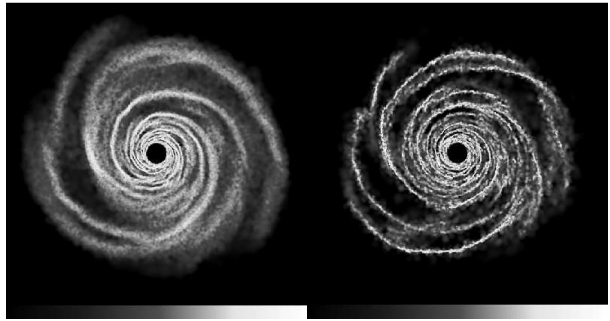,width=80.0mm}}
\caption{\small Surface density structure of particles embedded in a 
self-gravitating gas disk. a) The left-hand panel shows that
the distribution of 10 m radius particles is similar to that of the 
gas disk, because these particles are not influenced strongly 
by gas drag. b) The right-hand panel illustrates that 50 cm 
particles are strongly influenced by gas drag and become 
concentrated into the GI-spirals with density enhancements of 
an order of magnitude or more.
Figures adapted from {\it Rice et al.} (2004).} 
\label{dust}
\end{figure}

An analogous process could occur in a self-gravitating disk, where 
structures formed by GI activity, such as the centers of the 
spiral arms, are pressure and 
density maxima. In such a case, drag force results in solid particles
drifting towards the centers of these structures, with the most
significant effect occurring for those particles that would,
in a smooth, laminar disk, have the largest inward radial velocities. 

If disks around very young protostars do indeed undergo a self-gravitating
phase, then we would expect the resulting spiral structures to influence
the evolution of the solid particles in the disk ({\it Haghighipour and Boss},
2003a,b). A GI-active disk will also transport dust grains small 
enough to remain tied to the gas across distances of many AU's in only 
1,000 yrs or so ({\it Boss}, 2004b), a potentially important process for 
explaining the components of primitive meteorites (see the chapter by 
{\it Alexander et al.}). {\it Boley and Durisen} (2006) show that, 
in only one pass of a spiral shock, hydraulic jumps induced by shock 
heating can mix gas and entrained dust radially and vertically
over length-scales $\sim H$ through the generation of huge breaking waves.  
The presence of chondrules in primitive chondritic meteorites is 
circumstantial evidence that the Solar Nebula experienced a self-gravitating 
phase in which spiral shock waves provided the flash heating 
required to explain their existence ({\it Boss and Durisen}, 
2005a,b; {\it Boley et al.}, 2005). 

To test how a self-gravitating phase in a protostellar disk influences
the evolution of embedded particles, {\it Rice et al.} (2004) perform
3D self-gravitating disk simulations that include particles evolved 
under the influence of both disk self-gravity and gas drag. In their 
simulations, they consider both 10 m particles, which, for the 
chosen disk parameters, are only weakly coupled to the gas, 
and 50 cm particles that are significantly influenced by the gas drag.  
Fig. \ref{dust}a shows the surface density structure of the 10 m 
particles one outer rotation period after they were introduced into the gas disk. 
The structure in the particle disk matches closely that of the gas disk (not 
shown) showing that these particles are influenced by the gravitational force 
of the gas disk, but not so strongly influenced by gas drag. 
Fig. \ref{dust}b shows the surface density structure of 50 cm particles 
at the same epoch. Particles of this size are influenced by gas drag and
Fig. \ref{dust}b shows that, compared to the 10 m particles,
these particles become strongly concentrated into the GI-induced 
spiral structures. 

The ability of solid particles to become concentrated in the center
of GI-induced structures suggests that, even if giant planets do not form 
directly via GI's, a self-gravitating phase may still play an important
role in giant planet formation.  The solid particles may achieve
densities that could accelerate grain
growth either through an enhanced collision rate or through
direct gravitational collapse of the particle sub-disk ({\it Youdin and Shu}, 
2002). {\it Durisen et al.} (2005) also note that
dense rings can be formed near the boundaries between GI-active and
inactive regions of a disk (e.g., the central disk in Fig. \ref{teff}). 
Such rings are ideal sites for the concentration 
of solid particles by gas drag, possibly leading to accelerated growth of 
planetary embryos.  Even if processes like these do not contribute 
directly to planetesimal growth, GI's may act to prevent the loss of solids 
by migration toward the proto-Sun. The complex and time-variable structure 
of GI activity should increase the residence time of solids in the disk and 
potentially give them enough time to become sufficiently massive to 
decouple from the disk gas. 

\section{\textbf{PLANET FORMATION}}

The relatively high frequency ($\sim$10\%) of solar-type stars with giant 
planets that have orbital periods less than a few years suggests that 
longer-period planets may be quite frequent. Perhaps $\sim$ 12 to 25\% of G 
dwarfs may have gas giants orbiting within $\sim$10 AU. If so, gas giant 
planet formation must be a fairly efficient process. Because roughly half of
protoplanetary disks disappear within 3 Myr or less ({\it Bally et al.}, 1998; 
{\it Haisch et al.}, 2001; {\it Bric\~eno et al.}, 2001; 
{\it Eisner and Carpenter}, 2003), core accretion may not be able 
to produce a high frequency of gas giants. There is also now strong 
theoretical ({\it Yorke and Bodenheimer}, 1999) 
and observational ({\it Osorio et al.}, 2003; {\it Rodr\'iguez et al.}, 2005; 
{\it Eisner et al.}, 2005) evidence that disks around very young protostars 
should indeed be sufficiently massive to experience GI's. {\it Rodr\'iguez et al.} 
(2005) show a 7 mm VLA image of a disk around a Class 0 protostar that 
may have a mass half that of the central star. 

Hybrid scenarios may help 
remove the bottleneck by concentrating meter-sized solids, but it is not 
clear that they can shorten the overall time scale for core accretion,
which is limited by the time needed for the growth of 10 $M_\oplus$ 
cores and for accretion of a large gaseous envelope. {\it Durisen et al.}
(2005) suggest that the latter might be possible in dense rings, but 
detailed calculations of core growth or envelope accretion in the 
environment of a dense ring do not now exist. Disk instability, 
on the other hand, has no problem forming gas giants rapidly in 
even the shortest-lived protoplanetary disk. Most stars form in
regions of high-mass star formation ({\it Lada and Lada}, 2003) where 
disk lifetimes should be the shortest due to loss of outer disk gas by
UV irradiation.

There is currently disagreement about whether GI's are stronger 
in low-metallicity systems ({\it Cai et al.}, 2006) or whether their strength
is relatively insensitive to the opacity of the disk ({\it Boss}, 2002a). In 
either case, if disk instability is correct, we would expect that even 
low-metallicity stars could host gas giant planets. The growth of
cores in the core accretion mechanism is hastened by higher metallicity 
through the increase in surface density of solids ({\it Pollack et al.}, 1996), 
although the increased envelope opacity, which slows the collapse of
the atmosphere, works in the other direction ({\it Podolak}, 2003).  The
recent observation of a Saturn mass object, orbiting the metal-rich
star HD 149026, with a core mass equal to
approximately half the planet's mass ({\it Sato et al.}, 2005) has been
suggested as a strong confirmation of the core accretion model. 
It has, however, yet to be shown that the core accretion model can 
produce a core with such a relatively large mass. If this core was
produced by core accretion, it seems that it never achieved a 
runaway growth of its envelope; yet, in the case of Jupiter, the core
accretion scenario requires efficient accumulation of a massive 
envelope around a relatively low-mass core. 

The correlation of short-period gas giants with high metallicity 
stars is often interpreted as strong evidence in favor of
core accretion ({\it Laws et al.}, 2003; {\it Fischer et al.}, 
2004; {\it Santos et al.}, 2004). The {\it Santos et al.} (2004) 
analysis, however, shows that even the stars with the lowest metallicities 
have detectable planets with a frequency comparable to or higher than
that of the stars with intermediate metallicities. {\it Rice et al.} (2003c) have 
shown that the metallicity distribution of systems with at least 
one massive planet ($M_{pl} > 5 M_{Jup}$) on an eccentric orbits of 
moderate semi-major axis does not have the same metal-rich 
nature as the full sample of extrasolar planetary systems. 
Some of the metallicity correlation
can be explained by the observational bias of the spectroscopic method
in favor of detecting planets orbiting stars with strong metallic absorption 
lines. The residual velocity jitter typically increases from a few m/s for 
solar metallicity to 5 -- 16 m/s for stars with 1/4 the solar metallicity or less.
In terms of extrasolar planet search space, this could account
for as much as a factor of two difference in the total number of
planets detected by spectroscopy. A spectroscopic search of 98 stars
in the Hyades cluster, with a metallicity 35\% greater than solar, 
found nothing, whereas about 10 hot Jupiters should have been found,
assuming the same frequency as in the solar neighborhood 
({\it Paulson et al.}, 2004). 

{\it Jones} (2004) found that the average metallicity of planet-host stars 
increased from $\sim$ 0.07 to $\sim$ 0.24 dex for planets with semimajor 
axes of $\sim 2$ AU to $\sim 0.03$ AU, suggesting a trend toward 
shortest-period planets orbiting the most metal-rich stars. 
Similarly, {\it Sozzetti} (2004) showed that both metal-poor and metal-rich 
stars have increasing numbers of planets as the orbital period increases
but only the metal-rich stars have an excess of the shortest
period planets. This could imply that the metallicity correlation 
is caused by inward orbital migration, if low-metallicity stars have 
long-period giant planets that seldom migrate inward. 

 Lower disk metallicity results in slower Type II inward migration 
({\it Livio and Pringle}, 2003), the likely dominant mechanism for planet
migration (see the chapter by {\it Papaloizou et al.}). This is because
with increased metallicity, the disk viscosity $\nu$ increases. In 
standard viscous accretion disk theory (e.g., {\it Ruden and Pollack}, 1991) 
$\nu = \alpha c_s H$. Lower disk metallicity leads
to lower disk opacity, lower disk temperatures, lower sound speeds, 
and a thinner disk. As $\nu$ decreases with lowered metallicity, the 
time scale for Type II migration increases. {\it Ruden and Pollack} (1991) 
found that viscous disk evolution times increased by a factor of about 20
when $\nu$ decreased by a factor of 10. It remains to be seen if this effect
is large enough to explain the rest of the correlation. If disk instability is 
operative and if orbital migration is the major source of the metallicity 
correlation, then metal-poor stars should have planets on 
long-period orbits.

Disk instability may be necessary to account for the long-period giant 
planet in the M4 globular cluster ({\it Sigurdsson et al.}, 2003), where 
the metallicity is 1/20 to 1/30 solar metallicity. The absence of 
short-period Jupiters in the 47 Tuc globular cluster ({\it Gilliland 
et al.}, 2000) with 1/5 solar metallicity could be explained by the slow 
rate of inward migration due to the low metallicity. Furthermore, if 
47 Tuc initially contained OB stars, photoevaporation of the
outer disks may have occurred prior to inward orbital migration of any
giant planets, preventing their evolution into short-period planets,
though other factors (i.e., crowding) can also be important in these 
clusters.

The M dwarf GJ 876 is orbited by a pair of gas giants (as well as a much 
smaller mass planet) and other M dwarfs have giant planets as well ({\it Butler et al.},
2004), though apparently not as frequently as the G dwarfs. 
{\it Laughlin et al.} (2004) found that core accretion was too slow to 
form gas giants around M dwarfs because of the longer orbital periods. Disk 
instability does not a have similar problem for M dwarfs, and disk instability 
predicts that M, L, and T dwarfs should have giant planets.

With disk instability, one Jupiter mass of disk gas has at most 
$\sim 6 M_\oplus$ of elements suitable to form a rock/ice core.
The preferred models of the Jovian interior imply that Jupiter's core 
mass is less than $\sim 3 M_\oplus$ ({\it Saumon and Guillot}, 2004); 
Jupiter may even have no core at all. These models seem to be 
consistent with formation by disk instability and inconsistent with 
formation by core accretion, which requires a more massive core. As a 
result, the possibility of core erosion has been raised ({\it Saumon and 
Guillot}, 2004). If core erosion can occur, core masses may lose much 
of their usefulness as formation constraints.
Saturn's core mass appears to be larger than that of Jupiter 
({\it Saumon and Guillot}, 2004), perhaps $\sim 15 M_\oplus$,
in spite of it being the smaller gas giant. Core erosion would
only make Saturn's initial core even larger. Disk instability can
explain the larger Saturnian core mass ({\it Boss et al.}, 2002). 
Proto-Saturn may have started out with a mass larger than that of 
proto-Jupiter, but its excess gas may have been lost by UV 
photoevaporation, a process that could also form Uranus and Neptune.
Disk instability predicts that inner gas giants should be accompanied 
by outer ice giant planets in systems which formed in OB associations
due to strong UV photoevaporation. In low-mass star-forming regions, 
disk instability should produce only gas giants, without outer 
ice giants. 

Disk instability predicts that even the youngest stars should
show evidence of gas giant planets ({\it Boss}, 1998b),
whereas core accretion requires several Myr or more to form 
gas giants ({\it Inaba et al.}, 2003). 
A gas giant planet seems to be orbiting at $\sim10$ AU 
around the 1 Myr-old star CoKu Tau/4 ({\it Forrest et al.}, 2004), based on 
a spectral energy distribution showing an absence of disk dust 
inside 10 AU (for an alternative perspective, see {\it Tanaka et al.}, 2005). 
Several other 1 Myr-old stars show similar evidence for 
rapid formation of gas giant planets. The direct detection of a possible
gas giant planet around the 1 Myr-old star GQ Lup 
({\it Neuh\"auser et al.}, 2005) similarly requires a rapid planet 
formation process. 

We conclude that there are significant observational arguments to 
support the idea that disk instability, or perhaps a hybrid theory where 
core accretion is accelerated by GI's, might be required to form some 
if not all gas giant planets. Given the major uncertainties in the 
theories, 
observational tests will be crucial for determining the 
relative proportions of giant planets produced by the competing 
mechanisms.

\bigskip
\textbf{Acknowledgments.} {\small R.H.D.'s contribution was 
supported by NASA grants NAG5-11964 and NNG05GN11G and 
A.P.B.'s  by NASA grants NNG05GH30G, NNG05GL10G, and 
NCC2-1056. Support for A.F.N. was provided by the U.S. 
Department of Energy 
under contract W-7405-ENG-36, for which this is publication 
LA-UR--05-7851. We would like to thank S. Michael for 
invaluable assistance in manuscript preparation, an anonymous referee
for substantive improvements, and A.C. Mej\'ia,
A. Gawryszczak, and V. Springel for allowing us to premier their
comparison calculations in Section 4.4. FLASH was 
in part developed by the DOE-supported ASC/Alliance Center for 
Astrophysical Thermonuclear Flashes at the University of Chicago 
and was run on computers at Warsaw's Interdisciplinary Center for 
Mathematical and Computational Modeling.

\centerline\textbf{REFERENCES}
\bigskip
\parskip=0pt
{\small
\baselineskip=11pt

\refs
Adams F. C., Shu F. H., and Lada C. J. (1988)
{\em Astrophys. J., 326}, 865-883.

\refs 
Adams F. C., Ruden S. P., and Shu F. H. (1989)
{\em Astrophys. J., 347}, 959-976.

\refs 
Armitage P. J., Livio M., and Pringle J. E. (2001)
{\em Mon. Not. R. Astron. Soc., 324}, 705-711.

\refs 
Balbus S. A. and Papaloizou J. C. B. (1999)
{\em Astrophys. J., 521}, 650-658.

\refs
Balsara D. S. (1995) 
{\em J. Comput. Phys., 121}, 357-372.

\refs 
Bate M. R. and Burkert A. (1997) 
{\em Mon. Not. R. Astron. Soc., 228}, 1060-1072.

\refs 
Bally J., Testi L., Sargent A., and Carlstrom J. 
(1998) {\it Astron. J., 116}, 854-859. 

\refs 
Beckwith S. V. W., Sargent A. I., Chini R. S., and G\"usten R. (1990)
{\em Astron. J., 99}, 924-945.

\refs
Benz W. (1990) In {\em The Numerical Modeling of Nonlinear Stellar
Pulsations} (J. R. Buchler, ed.), pp. 269-288. Kluwer, Boston.

\refs  
Boffin H. M. J., Watkins S. J., Bhattal A. S., Francis N., 
and Whitworth A. P. (1998) 
{\em Mon. Not. R. Astron. Soc., 300}, 1189-1204.

\refs
Bodenheimer P. and Pollack J. B. (1986)
{\em Icarus, 67}, 391-408.

\refs 
Bodenheimer P., Yorke H. W., R\'o\.zyczka M., and Tohline J. E. (1990)
{\em Astrophys. J., 355}, 651-660.

\refs 
Boley A. C. and Durisen R. H. (2006)
{\em Astrophys. J.}, in press (astro-ph 0510305).

\refs 
Boley A. C., Durisen R. H., and Pickett M. K. (2005) 
In {\em Chondrites and the Protoplanetary Disk} 
(A. N. Krot et al., eds.), pp. 839-848. 
ASP Conference Series, San Francisco.

\refs 
Boss A. P. (1997) 
{\em Science, 276}, 1836-1839.

\refs 
Boss A. P. (1998a) 
{\em Astrophys. J., 503}, 923-937.

\refs 
Boss A. P. 
(1998b) {\em Nature, 395}, 141-143. 

\refs 
Boss A. P. (2000)
{\em Astrophys. J., 536}, L101-L104. 

\refs 
Boss A. P. (2001) 
{\em Astrophys. J., 563}, 367-373.

\refs 
Boss A. P. (2002a) 
{\em Astrophys. J., 567}, L149-L153.

\refs 
Boss A. P. (2002b) 
{\em Astrophys. J., 576}, 462-472.

\refs
Boss A. P. (2002c)
{\em Earth Planet. Sci. Let., 202}, 513-523.

\refs 
Boss A. P. (2003) 
{\em Lunar Planet. Inst., 34}, 1075-1076.

\refs 
Boss, A. P. (2004a) 
{\em Astrophys. J., 610}, 456-463.

\refs 
Boss A. P. (2004b) 
{\em Astrophys. J., 616}, 1265-1277.

\refs 
Boss A. P. (2005) 
{\em Astrophys. J., 629}, 535-548.

\refs 
Boss A. P. (2006)
{\em Astrophys. J.}, in press.

\refs 
Boss A. P. and Durisen R. H. (2005a) 
{\em Astrophys. J., 621}, L137-L140.

\refs 
Boss A. P. and Durisen R. H. (2005b)
In {\em Chondrites and the Protoplanetary Disk} 
(A. N. Krot et al., eds.), pp. 821-838.
ASP Conference Series, San Francisco.

\refs
Boss A. P. and Myhill E. A. (1992) 
{\em Astrophys. J. Suppl., 83}, 311-327.

\refs
Boss A. P. and Yorke H. W. (1996)
{\em Astrophys. J., 496}, 366-372.

\refs 
Boss A. P., Wetherill G. W., and Haghighipour N. (2002) 
{\em Icarus, 156}, 291-295.

\refs 
Brice\~no C., Vivas A. K., Calvet N., Hartmann L., 
Pachecci R. et al. (2001) 
{\em Science, 291}, 93-96. 

\refs 
Butler R. P., Vogt S. S., Marcy G. W., Fischer D. A., 
Wright J. T. et al. 
(2004) {\em Astrophys. J., 617}, 580-588.

\refs
Bryden G., Lin D. N. C., and Ida S. (2000) 
{\em Astrophys. J., 544}, 481-495.

\refs 
Cai K., Durisen R. H., Michael S., Boley A. C., 
Mej\'ia A. C., Pickett M. K., and D'Alessio P. (2006) 
{\em Astrophys. J., 636}, L149-L152.

\refs 
Cameron A. G. W. (1978) 
{\em Moon Planets, 18}, 5-40.

\refs 
Chiang E. I. and Goldreich P. (1997)  
{\em Astrophys. J., 490}, 368-376.

\refs
Colella P. and Woodward P. R. (1984) 
{\em J. Comp. Phys., 54}, 174-201.

\refs
Cuzzi J. N., Hogan R. C., Paque J. M., and Dobrovolskis A. R. 
(2001) {\em Astrophys. J., 546}, 496-508.

\refs 
D'Alessio P., Calvet N., and Hartmann L. (1997) 
{\em Astrophys. J., 474}, 397-406.  

\refs 
D'Alessio P., Cant\'o J., Calvet N., and Lizano S. (1998) 
{\em Astrophys. J., 500}, 411-427.

\refs
Durisen R. H. (2006)
In {\em A Decade of Extrasolar Planets Around Normal
Stars} (M. Livio, ed.), in press. University
Press, Cambridge. 

\refs 
Durisen R. H., Cai K., Mej\'ia A. C., and Pickett M. K. (2005) 
{\em Icarus, 173}, 417-424.

\refs 
Durisen R. H., Mej\'ia A. C., and Pickett B. K. (2003)
{\em Rec. Devel. Astrophys., 1}, 173-201.

\refs 
Durisen R. H., Mej\'ia A. C., Pickett B. K., and Hartquist T. W. (2001) 
{\em Astrophys. J., 563}, L157-L160.

\refs 
Eisner J. A. and Carpenter J. M. (2003) 
{\em Astrophys. J., 598}, 1341-1349. 

\refs 
Eisner J. A., Hillenbrand L. A., Carpenter J. M., and Wolf S. (2005)
{\em Astrophys. J., 635}, 396-421.

\refs
Fischer D., Valenti J. A. and Marcy G. (2004) 
In {\em IAU Symposium \#219: Stars as Suns: Activity, Evolution, and
Planets} (A. K. Dupree and A. O. Benz, eds.), pp. 29-38.
APS Conference Series, San Francisco.

\refs 
Forrest W. J., Sargent B.,  Furlan E., D'Alessio P., Calvet N. et al.
(2004) {\em Astrophys. J. Suppl., 154}, 443-447.

\refs 
Fleming T. and Stone J. M. (2003)
{\em Astrophys. J., 585}, 908-920.

\refs
Fryxell B., Arnett D., and M\"uller E. (1991) 
{\em Astrophys. J., 367}, 619-634. 

\refs
Fryxell B.,  Olson K., Ricker P., Timmes F. X., Zingale M. et al. (2000)
{\em Astrophys. J. Suppl., 131}, 273-334.

\refs 
Gammie C. F. (1996) 
{\em Astrophys. J., 457}, 355-362. 

\refs 
Gammie C. F. (2001) 
{\em Astrophys. J., 553}, 174-183.

\refs 
Gilliland R. L., Brown T. M., Guhathakurta P., Sarajedini A., 
Milone E. F. et al. (2000) 
{\it Astrophys. J., 545}, L47-L51. 

\refs 
Goldreich P. and Lynden-Bell D. (1965)
{\em Mon. Not. R. Astron. Soc., 130}, 125-158.

\refs 
Haghighipour N. and Boss A. P. (2003a) 
{\em Astrophys. J., 583}, 996-1003.

\refs 
Haghighipour N. and Boss A. P. (2003b) 
{\em Astrophys. J., 598}, 1301-1311.

\refs 
Haisch K. E., Lada E. A., and Lada C. J. (2001) 
{\it Astrophys. J., 553}, L153-L156. 

\refs 
Inaba S., Wetherill G. W., and Ikoma M. (2003) 
{\it Icarus, 166}, 46-62.

\refs 
Johnson B. M. and Gammie C. F. (2003)
{\em Astrophys. J., 597}, 131-141.

\refs 
Johnson B. M. and Gammie C. F. (2006)
{\em Astrophys. J., 636}, 63-74.

\refs 
Johnstone D., Hollenbach D., and Bally J. (1998)
{\em Astrophys. J., 499}, 758-776.

\refs
Jones H. R. A. (2004) 
In {\it The Search for Other Worlds: Fourteenth Astrophysics Conference,
AIP Conference Proceedings, 713}, pp. 17-26. AIP Conference Proceedings,
New York.

\refs
Klahr H. H. (2003) 
In {\it Scientific Frontiers in
Research on Extrasolar Planets} (D. Deming and S. Seager, eds.), 
pp. 277-280. ASP Conference Series, San Francisco.

\refs 
Klahr H. H. and Bodenheimer P. (2003)
{\em Astrophys. J., 582}, 869-892.

\refs 
Klahr H. H. and Henning T. (1997)
{\em Icarus, 128}, 213-229.

\refs
Kley W. (1999) 
{\em Mon. Not. R. Astron. Soc., 303}, 696-710.

\refs
Koller J., Li H., and Lin D. N. C. (2003) 
{\em Astrophys. J., 596}, L91-94.

\refs 
Kuiper G. P. (1951)
In {\em Proceedings of a Topical Symposium} 
(J. A. Hynek, ed.), pp. 357-424. McGraw-Hill, New York.

\refs 
Lada C. J. and Lada E. A. (2003) 
{\it Ann. Rev. Astron. Astrophys., 41}, 57-115.

\refs
Larson R. B. (1984)
{\em Mon. Not. R. Astron. Soc., 206}, 197-207. 

\refs 
Laughlin G. and Bodenheimer P. (1994) 
{\em Astrophys. J., 436}, 335-354.  

\refs 
Laughlin G. and R\'o\.zyczka M. (1996)
{\em Astrophys. J., 456}, 279-291.

\refs 
Laughlin G., Korchagin V., and Adams F. C. (1997)
{\em Astrophys. J., 477}, 410-423.

\refs 
Laughlin G., Korchagin V., and Adams F. C. (1998)
{\em Astrophys. J., 504}, 945-966.

\refs 
Laughlin G., Bodenheimer P., and Adams F. C. (2004) 
{\it Astrophys. J., 612}, L73-L76.

\refs
Laws C., Gonzalez G., Walker K. M., Tyagi S., Dodsworth J. et al.
(2003) {\it Astron. J., 125}, 2664-2677.

\refs
Lin D. N. C. and Papaloizou J. C. B. (1993) 
In {\em Protostars and Planets III} 
(E. H. Levy and J. I. Lunine, eds.), pp. 749-835.
Univ. of Arizona, Tucson.

\refs 
Lin D. N. C. and Pringle J. E. (1987) 
{\em Mon. Not. R. Astron. Soc., 225}, 607-613.  

\refs 
Lin D. N. C., Laughlin G., Bodenheimer P., 
and R\'o\.zyczka M. (1998)
{\em Science, 281}, 2025-2027.

\refs 
Livio M. and Pringle J. E. (2003) 
{\it Mon. Not. R. Astron. Soc., 346}, L42-L44. 

\refs 
Lodato G. and Rice W. K. M. (2004)
{\em Mon. Not. R. Astron. Soc., 351}, 630-642.

\refs 
Lodato G. and Rice W. K. M. (2005)
{\em Mon. Not. R. Astron. Soc., 358}, 1489-1500.

\refs 
Lubow S. H. and Ogilvie G. I. (1998)
{\em Astrophys. J., 504}, 983-995.

\refs 
Lubow S. H., Siebert M., and Artymowicz P. (1999) 
{\em Astrophys. J., 526}, 1001-1012.

\refs
Lufkin G., Quinn T. Wadsley J., Stadel J., and Governato F. (2004)
{\em Mon. Not. R. Astron. Soc., 347}, 421-429.

\refs 
Mayer L., Quinn T., Wadsley J., and Stadel J. (2002)
{\em Science 298}, 1756-1759.

\refs 
Mayer L., Quinn T., Wadsley J., and Stadel J. (2004a)
{\em Astrophys. J., 609}, 1045-1064.

\refs 
Mayer L., Wadsley J., Quinn T., and Stadel J. (2004b) 
In {\em Extrasolar Planets: Today and Tomorrow} 
(J.-P. Beaulieu et al., eds.), pp. 290-297. 
ASP Conference Series, San Francisco.

\refs
Mayer L., Wadsley J., Quinn T., and Stadel J. (2005)
{\em Mon. Not. R. Astron. Soc., 363}, 641-648.

\refs 
Mej\'ia A. C. (2004) 
Ph.D. dissertation, Indiana University.

\refs 
Mej\'ia A. C., Durisen R. H., Pickett M. K., and Cai K. (2005)
{\em Astrophys. J., 619}, 1098-1113.

\refs
Mihalas D. (1977) 
{\em Stellar Atmospheres}. Univ. of
Chicago, Chicago.

\refs
Monaghan J. J. (1992) 
{\em Ann. Rev. Astron. Astrophys., 30}, 543-574.

\refs
Myhill E. A. and Boss A. P. (1993)
{\em Astrophys. J. Suppl.,  89}, 345-359.

\refs 
Nelson A. F. (2000)
{\em Astrophys. J., 537}, L65-L69.

\refs
Nelson A. F. (2006)
{\em Mon. Not. R. Astron. Soc.}, submitted.

\refs
Nelson A. F. and Benz W. (2003) 
{\em Astrophys. J., 589}, 578-604.

\refs 
Nelson A. F., Benz W., Adams F. C., and Arnett D. (1998)
{\em Astrophys. J., 502}, 342-371.

\refs 
Nelson A. F., Benz W., and Ruzmaikina T. V. (2000)
{\em Astrophys. J., 529}, 357-390.

\refs
Nelson R. P., Papaloizou J. C. B., Masset F., and Kley W.
(2000) {\em Mon. Not. R. Astron. Soc., 318}, 18-36.

\refs 
Neuh\"auser R., Guenther E. W., Wuchterl G., Mugrauer M.,
Bedalov A., and Hauschildt P. H. (2005) 
{\it Astron. Astrophys., 435}, L13-L16.

\refs 
Osorio M., D'Alessio P., Muzerolle J., Calvet N., and Hartmann L. (2003) 
{\em Astrophys. J., 586}, 1148-1161.

\refs 
Paczy\'nski B. (1978)
{\em Acta Astron., 28}, 91-109.

\refs 
Papaloizou J. C. B. and Savonije G. (1991)
{\em Mon. Not. R. Astron. Soc., 248}, 353-369.

\refs
Paulson D. B., Saar S. H., Cochran W. D., and Henry G. W. (2004) 
{\it Astron. J., 127}, 1644-1652.

\refs
Pickett B. K. (1995) 
Ph.D. dissertation, Indiana University.

\refs
Pickett B. K., Durisen R. H., and Davis G. A. (1996)
{\em Astrophys. J., 458}, 714-738.

\refs
Pickett B. K., Cassen P., Durisen R. H., and Link R. P. (1998)
{\em Astrophys. J., 504}, 468-491.

\refs 
Pickett B. K., Cassen P., Durisen R. H., and Link R. P. (2000a)
{\em Astrophys. J., 529}, 1034-1053.

\refs
Pickett B. K., Durisen R. H., Cassen P., and Mej\'ia A. C. (2000b)
{\em Astrophys. J., 540}, L95-98. 

\refs 
Pickett B. K., Mej\'ia A. C., Durisen R. H., Cassen P. M., 
Berry D. K., and Link R. P. (2003) 
{\em Astrophys. J., 590}, 1060-1080.

\refs
Podolak M. (2003) 
{\it Icarus, 165}, 428-437.

\refs 
Pollack J. B., Hubickyj O., Bodenheimer P., Lissauer J. J., 
Podolak M., and Greenzweig Y. (1996) 
{\em Icarus, 124}, 62-85. 

\refs
Pringle J. E. (1981) 
{\it Ann. Rev. Astron. Astrophys., 19}, 137-162.

\refs 
Rafikov R. R. (2005)
{\em Astrophys. J., 621}, L69-L72.

\refs 
Rice W. K. M., Armitage P. J., Bate M. R., and Bonnel I. A. (2003a)
{\em Mon. Not. R. Astron. Soc., 338}, 227-232.

\refs 
Rice W. K. M., Armitage P. J., Bate M. R., and Bonnell I. A. (2003b) 
{\em Mon. Not. R. Astron. Soc., 339}, 1025-1030.

\refs 
Rice W. K. M., Armitage P. J., Bate M. R. and Bonnell I. A. (2003c) 
{\it Mon. Not. R. Astron. Soc., 346}, L36-L40.

\refs 
Rice W. K. M., Lodato G., and Armitage P. J. (2005) 
{\em Mon. Not. R. Astron. Soc., 364}, L56-L60. 

\refs 
Rice W. K. M., Lodato G., Pringle J. E., Armitage P. J., 
and Bonnell I. A. (2004) 
{\em Mon. Not. R. Astron. Soc., 355}, 543-552.

\refs 
Rodr\'iguez L. F., Loinard L., D'Alessio P., Wilner D. J., and
Ho P. T. P. (2005)
{\em Astrophys. J., 621}, 133-136. 

\refs
Ruden S. P. and Pollack J. B. (1991) 
{\em Astrophys. J., 375}, 740-760.

\refs
Santos N. C., Israelian G., and Mayor M. (2004) 
{\em Astron. Astrophys., 415}, 1153-1166.

\refs
Sato B., Fischer D. A., Henry G. W., Laughlin G., 
Butler R. P. et al. (2005) 
{\em Astrophys. J., 633}, 465-473. 

\refs 
Saumon D. and Guillot T. (2004) 
{\em Astrophys. J., 609}, 1170-1180.

\refs 
Shu F. H., Tremaine S., Adams F. C., and Ruden S. P. (1990)
{\em Astrophys. J., 358}, 495-514.

\refs 
Sigurdsson S., Richer H. B., Hansen B. M., Stairs I. H., 
and Thorsett S. E. (2003) 
{\it Science, 301}, 193-196.

\refs
Springel V. (2005) 
{\em Mon. Not. R. Astron. Soc., 364}, 1105-1134. 

\refs
Springel V., Yoshida N., and White S. D. M. (2001) 
{\em New Astron., 6}, 79-117.

\refs 
Sozzetti A. (2004) 
{\it Mon. Not. R. Astron. Soc., 354}, 1194-1200.

\refs
Stone J. M. and Norman M. L. (1992) 
{\em Astrophys. J. Suppl., 80}, 753-790.

\refs 
Tanaka H., Himeno Y., and Ida S. (2005) 
{\em Astrophys. J., 625}, 414-426.

\refs 
Tohline J. E. (1980) 
{\em Astrophys. J., 235}, 866-881.

\refs 
Toomre A. (1964)
{\em Astrophys. J., 139}, 1217-1238.

\refs 
Tomley L., Cassen P., and Steiman-Cameron T. Y. (1991)
{\em Astrophys. J., 382}, 530-543.

\refs 
Truelove J. K., Klein R. I., McKee C. F., Holliman J. H. II, 
Howell L. H., and Greenough J. A. (1997) 
{\em Astrophys. J., 489}, L179-L183.

\refs
Wadsley J., Stadel J., and Quinn T. (2004)
{\em New Astron., 9}, 137-158.

\refs 
Weidenschilling S. J. (1977)
{\em Mon. Not. R. Astron. Soc., 180}, 57-70.

\refs 
Weidenschilling S. J. and Cuzzi J. N. (1993)
In {\em Protostars and Planets III}
(E. H. Levy and J. I. Lunine, eds.), pp. 1031-1060.
Univ. of Arizona, Tucson.

\refs 
Yorke H. W. and Bodenheimer P. (1999)
{\em Astrophys. J., 525}, 330-342.

\refs 
Youdin A. N. and Shu F. H. (2002)
{\em Astrophys. J., 580}, 494-505.
 
}

\end{document}